\documentclass[twocolumn,amsmath,amssymb]{revtex4}
\usepackage{color}

\usepackage{natbib,graphicx,amsmath}

\newcommand{\bea}{\begin{eqnarray}}
\newcommand{\eea}{\end{eqnarray}}

\begin{document}
\title{Axion Pyramid}
\author{Chan-Gyung Park${}^{1}$, Dahee Lee${}^{2}$,  Hyerim Noh${}^{3}$, Jai-chan Hwang${}^{4}$}
\address{${}^{1}$Division of Science Education and Institute of Fusion Science, Jeonbuk National University, Jeonju 54896, Republic of Korea
         \\
         ${}^{2}$Department of Earth Sciences, Chosun University, Jeonnam-Gwangju 61452, Republic of Korea
         \\
         ${}^{3}$Theoretical Astrophysics Group, Korea Astronomy and Space Science Institute, Daejeon, 34055, Republic of Korea
         \\
         ${}^{4}$Particle Theory and Cosmology Group, Center for Theoretical Physics of the Universe, Institute for Basic Science (IBS), Daejeon, 34051, Republic of Korea
         }


\begin{abstract}

A pronounced pyramid-shaped enhancement in the axion density power spectrum induced by the QCD phase transition is identified. The enhancement occurs near the horizon scale at the transition, with a characteristic comoving length scale comparable to the size of the Solar System, and amplifies the axion density perturbation by factors exceeding $3.5 \times 10^{13}$ relative to the CDM case. Under the hilltop initial condition, coupling to radiation triggers this enhancement. The initial condition is related to the axion mass, suggesting that observation of the axion pyramid in cosmology may constrain the QCD axion mass.

\end{abstract}
\maketitle

\section{Introduction}

The QCD axion in a coherent oscillation stage at the bottom of the potential is a strong candidate for the cold dark matter (CDM) \cite{Preskill-1983, Abbott-1983, Dine-1983, Khlopov-1985, Bianchi-1990, Sikivie-Yang-2009, Hwang-2009, Noh-2017}. During coherent oscillation with $\hbar H/(mc^2) \ll 1$, the axion as a massive scalar field is non-relativistic and can be described by Schr\"odinger's equation. The corresponding case was studied by Madelung \cite{Madelung-1927} in the early development of quantum mechanics. The Schr\"odinger equation in a polar decomposition leads to the well-known hydrodynamic continuity and momentum conservation equations with a curious form of quantum stress.

The quantum nature of the stress is supported by the fact that the term involves $\hbar^2$, and it accounts for all quantum mechanical phenomena such as the uncertainty principle, tunnelling, interference, and quantized energy levels, etc., when applied on atomic scales \cite{Bohm-1952}; in Bohm's case, the Schr\"odinger equation in polar decomposition leads to the continuity and Hamilton-Jacobi equations with an esoteric form of quantum potential. The nature of the stress, as opposed to pressure (isotropic stress), is clarified in \cite{Takabayasi-1952}. 

For the QCD axion, the quantum stress gives rise to a quantum Jeans scale, suppressing the density fluctuations below that scale. Its current physical size is about the size of the Solar-System, and is therefore negligible in cosmology. This physical size corresponds to a comoving scale comparable to the horizon size at the time of the QCD phase transition. The QCD phase transition is also known to cause an enhancement of density fluctuations near these scales. In this case, the implication for cosmology, with abundant compact axion DM structures on that scale, may differ significantly.

Various enhancement mechanisms for the axion density power spectrum during the QCD phase transition have been reported in the literature \cite{Hogan-1988, Kolb-1993, Kolb-1994, Kolb-1996, Schmid-1999, Eroncil-2023, Ayad-Schwarz-2025, Arvanitaki-2020, Sikivie-Xue-2022, Kitajima-2022}. The radiation coupling due to the temperature dependence of the axion mass during the phase transition and the self-coupling due to interaction terms in the axion potential can cause an enhancement of the power spectrum around the horizon scale during the phase transition \cite{Arvanitaki-2020, Kitajima-2022}. 

Previous studies \cite{Arvanitaki-2020, Kitajima-2022}, however, have not followed the numerical evolution far enough to cover the axion mass generation process. Due to the increasing frequency of rapid oscillations of the axion field, direct numerical integration cannot accurately capture the QCD phase transition process to its end. To numerically handle the rapid oscillation of the axion field, we develop a hierarchy of WKB approximations, from an exact treatment to increasingly approximate ones. 

We show that a conspicuous pyramid-shaped enhancement of the axion density power spectrum---reaching a colossal factor of $10^{27}$ or more depending on the initial location of the field or the axion mass---appears in a radiation-coupled model with hilltop initial conditions (ICs), see Fig.\ \ref{P}. For a bottom IC, the radiation coupling produces a bluer power spectrum than that in the CDM case, providing the basis for the huge enhancement induced by self-coupling. Even in the absence of radiation-coupling, the hilltop ICs cause huge enhancement in the power spectrum by factors exceeding $10^9$ relative to the CDM case, see Fig.\ \ref{P-q=0}. These enhancements had previously been noticed in direct numerical calculations \cite{Arvanitaki-2020, Kitajima-2022} which did not cover the entire phase transition process. Their precise shapes and magnitudes are now revealed in this work. 

On the other hand, for a massive axion without radiation coupling, we show a strong suppression of the axion density power spectrum compared with the CDM case below the horizon crossing scale at the time of the phase transition. The suppression is not surprising considering the quantum stress of the axion field, but the clarification of its origin requires a fluid formulation \cite{Park-2012}. 

As the QCD axion, we consider a scalar field with a radiation-coupled sine-Gordon potential \cite{Kitajima-2022}
\bea
   & & V (\phi, \varrho_r)
       = f^2 m_0^2 X (\varrho_r) [ 1 - \cos{(\phi/f)} ],
   \nonumber \\
   & & X (\varrho_r)
       \equiv {1 \over 1 + (\varrho_r / \varrho_{rc})^q},
\eea
where $[1 - \cos{(\phi/f)]}$ indicates the self-interaction; we set $c \equiv 1 \equiv \hbar$. The $X$-term indicates the radiation coupling due to the temperature-dependent mass $m \equiv m_0 \sqrt{X}$. For the QCD phase transition we take $q = 2.05$ and $T_{rc} = 0.15{\rm GeV}$ with $\varrho_r = {1 \over 2} g_* (T) \varrho_\gamma$ and the entropy density $s = (\varrho_r + p_r)/T = {2 \pi^2 \over 45} g_{*s} (T) T^3$ for the radiation and $\varrho_\gamma = {\pi^2 \over 15} T^4$ for photons \cite{Turner-1986, Saikawa-2018}; for $g_{*s} = g_*$ we have $p_r = {1 \over 3} \varrho_r$. We set $m_0 = 5.8 \times 10^{-6} {\rm eV} ( 10^{12} {\rm GeV}/f )$; without the radiation coupling, thus $q = 0$, the axion mass is $m = m_0/\sqrt{2}$. 

We consider a flat Friedmann cosmology with linear order scalar perturbations. Our metric convention is
\bea
   g_{00} = - \left( 1 + 2 \alpha \right), \;\;
       g_{0i} = - \chi_{,i}, \;\; g_{ij} = a^2 \left( 1 + 2 \varphi \right) \delta_{ij},
   \label{metric}
\eea
where $a$ is the cosmic scale factor, and $\alpha$, $\chi$ and $\varphi$ are linear-order perturbations without the temporal gauge condition imposed yet; $x^0 = t$. Algebraic details required for the numerical calculations are presented in the Appendix.

\section{Background evolution}

For the background, we have
\bea
   & & H^2 = {8 \pi G \over 3} \varrho + {\Lambda \over 3}, \quad
       \ddot \phi + 3 H \dot \phi + V_{,\phi} = 0,
   \nonumber \\
   & & \varrho = \varrho_r + \varrho_\phi, \quad
       p = {1 \over 3} \varrho_r + p_\phi,
   \nonumber \\
   & & \varrho_\phi \equiv {1 \over 2} \dot \phi^2 + V, \quad
       p_\phi \equiv {1 \over 2} \dot \phi^2 - V,
   \label{fluid-BG}
\eea
where $H \equiv \dot a/a$ and an overdot indicates a time derivative with $t$. In the early universe, up to the QCD phase transition, radiation dominated over other components. In the energy-momentum tensor we include the axion and radiation (including photons, massless neutrinos with an effective number of species $N_\nu = 3.04$, and other relativistic species excited during the QCD era). 

If we assume that $p_r = {1 \over 3} \varrho_r$ holds throughout the evolution, we then have $\varrho_r \propto a^{-4}$ and $a \propto t^{1/2}$; this assumption, which was also made in \cite{Arvanitaki-2020, Kitajima-2022}, is required for Eqs.\ (\ref{ddot-delta-RCG})-(\ref{EOM-pert-RCG}). For time-dependent $g_*$ and $g_{*s}$ we no longer have $p_r = {1 \over 3} \varrho_r$, and Eqs.\ (\ref{ddot-delta-eq-w})-(\ref{alpha-kappa}) must instead be solved directly. In the latter case, for the radiation density and pressure we use the tabulated energy-density and entropy degrees of freedom, $g_{*}(T)$ and $g_{*s}(T)$, provided in \cite{Saikawa-2018}. We determine $T(a)$ from entropy conservation and obtain the equation-of-state parameter $w \equiv p/\varrho$ and the sound speed $c_s^2 \equiv \dot p/\dot \varrho$ using
\bea
   w = {4 \over 3} {g_{*s} \over g_{*}} - 1, \quad
       c_s^2 = \Big( 3 + {d\ln g_{*s} \over d\ln T} \Big)^{-1}.
   \label{w-c_s}
\eea
For $p_r = {1 \over 3} \varrho_r$, thus $g_*=\textrm{constant}$, the scale factor at the QCD transition epoch ($T=T_{rc}$) is $a_c=1.565\times 10^{-12} a_0$, whereas with time-dependent $g_*$ and $g_{*s}$ it is $a_c=8.403\times 10^{-13} a_0$. The comoving wavenumber of the mode that crosses the Hubble horizon at $a_c$ is $k_c=(a_c H_c)/c = (1.498 \times 10^5 {\rm AU})^{-1}$ for $p_r=\frac{1}{3}\varrho_r$ and $k_c
=(0.985 \times 10^5 {\rm AU})^{-1}$ for time-dependent $g_{*}$ and $g_{*s}$.

In the early universe with $H$ larger than $m$, $\phi$ slowly rolls down the potential. As $H$ becomes smaller than $m$ with $\phi/f \ll 1$, which happens before the QCD phase transition, $\phi$ starts to oscillate with a frequency of order $m$. The equation of motion in Eq.\ (\ref{fluid-BG}) becomes
\bea
   \ddot \phi + 3 H \dot \phi + m^2 \phi = 0.
\eea
In the oscillating regime of the axion field, the oscillation timescale soon becomes much shorter than the integration timescale in $\ln{a}$ and we cannot directly solve the background evolution numerically. 

In this rapidly oscillating stage in the radiation-dominated era we consider the WKB solution \cite{Sikivie-Xue-2022}
\bea
   \phi = {\overline A \over \sqrt{m} a^{3/2}} 
        \cos{(\overline \Omega + \overline B )}, \quad
        \overline \Omega \equiv \int^t_{t_{\rm tr}} m d t.
   \label{phi-WKB-sol}
\eea
The amplitude $\overline A$ and phase $\overline B$ are constants determined by matching $\phi$ and $\dot \phi$ to their numerical values at the transition epoch $t_{\rm tr}$. For $H/m \ll 1$ we have
\bea
   \varrho_\phi = {m \overline A^2 \over 2 a^3}, \quad
       p_\phi = - \cos{(2 \overline \Omega + 2 \overline B)} \varrho_\phi.
\eea
Ignoring the rapid oscillation in the pressure by time averaging, we have $p_\phi = 0$, and the background axion behaves as a dust fluid or as CDM \cite{Preskill-1983, Abbott-1983, Dine-1983}. 

The axion energy density $\varrho_\phi$ at the end of the integration, {\it after} completion of the QCD phase transition, is normalized to reproduce the expected CDM density corresponding to the present epoch density parameter $\Omega_{\rm CDM} = 0.26$ \cite{Planck-2020}. This procedure imposes a strong correlation between the axion's initial position and its mass---the $f $ versus $\phi_i/f$ relation \cite{Turner-1986, Bae-2008, Visinelli-2009, Wantz-2010, Saikawa-2018, Arvanitaki-2020, Kitajima-2022}.

\section{Perturbation evolution}

To derive the axion-density power spectrum we consider the axion, radiation, and the scalar-type metric perturbations to linear order. Assuming the radiation-dominated environment during the QCD phase transition, we can solve the axion perturbation equation driven by {\it given} radiation and metric perturbations in the radiation comoving gauge setting the perturbed velocity of radiation $v_r \equiv 0$. 

In the purely radiation-dominated era, we have \cite{Bardeen-1980}
\bea
   & & \ddot \delta_r + H \dot \delta_r
       - \Big( 2 H^2 + {\Delta \over 3 a^2} \Big)
       \delta_r = 0,
   \label{ddot-delta-RCG}
\eea
where $\delta_r \equiv \delta \varrho_r/\varrho_r$. This follows from Eqs.\ (\ref{eq6-r}), (\ref{eq7-r}) and (\ref{eq4}). The solution, with $\Delta \equiv - k^2$, is \cite{Bardeen-1980}
\bea
   & & \delta_{r}
       = 4 C \Big( {\sin{x} \over x} - \cos{x} \Big),
   \label{delta_r-sol}
\eea 
where $x \equiv k \eta / \sqrt{3}$, and $\eta$ is the conformal time defined by $dt \equiv ad\eta$. We consider the axion perturbation in this radiation-dominated universe.

For the axion perturbation, we solve the equation of motion for $\delta \phi$
\bea
   & & \delta \ddot \phi + 3 H \delta \dot \phi
       - {\Delta \over a^2} \delta \phi
       + V_{,\phi\phi} \delta \phi
   \nonumber \\
   & & \qquad
       = {1 \over 2} \dot \phi \dot \delta_r
       + \Big( {1 \over 2} V_{,\phi}
       - V_{,\phi\varrho_r} \varrho_r \Big) \delta _r 
       \equiv H^2 S,
   \label{EOM-pert-RCG}
\eea
which follows from Eqs.\ (\ref{eq6-r})-(\ref{EOM-pert}). We can numerically integrate Eq.\ (\ref{EOM-pert-RCG}) for $\delta \phi$ using $\delta_r$ given in Eq.\ (\ref{delta_r-sol}) and using $\phi$ and $\dot \phi$ obtained by solving the second equation in Eq.\ (\ref{fluid-BG}).

\begin{figure}[t]
\hspace{-0.2cm} \vspace{.0cm}
\includegraphics[width=83mm,angle=0]{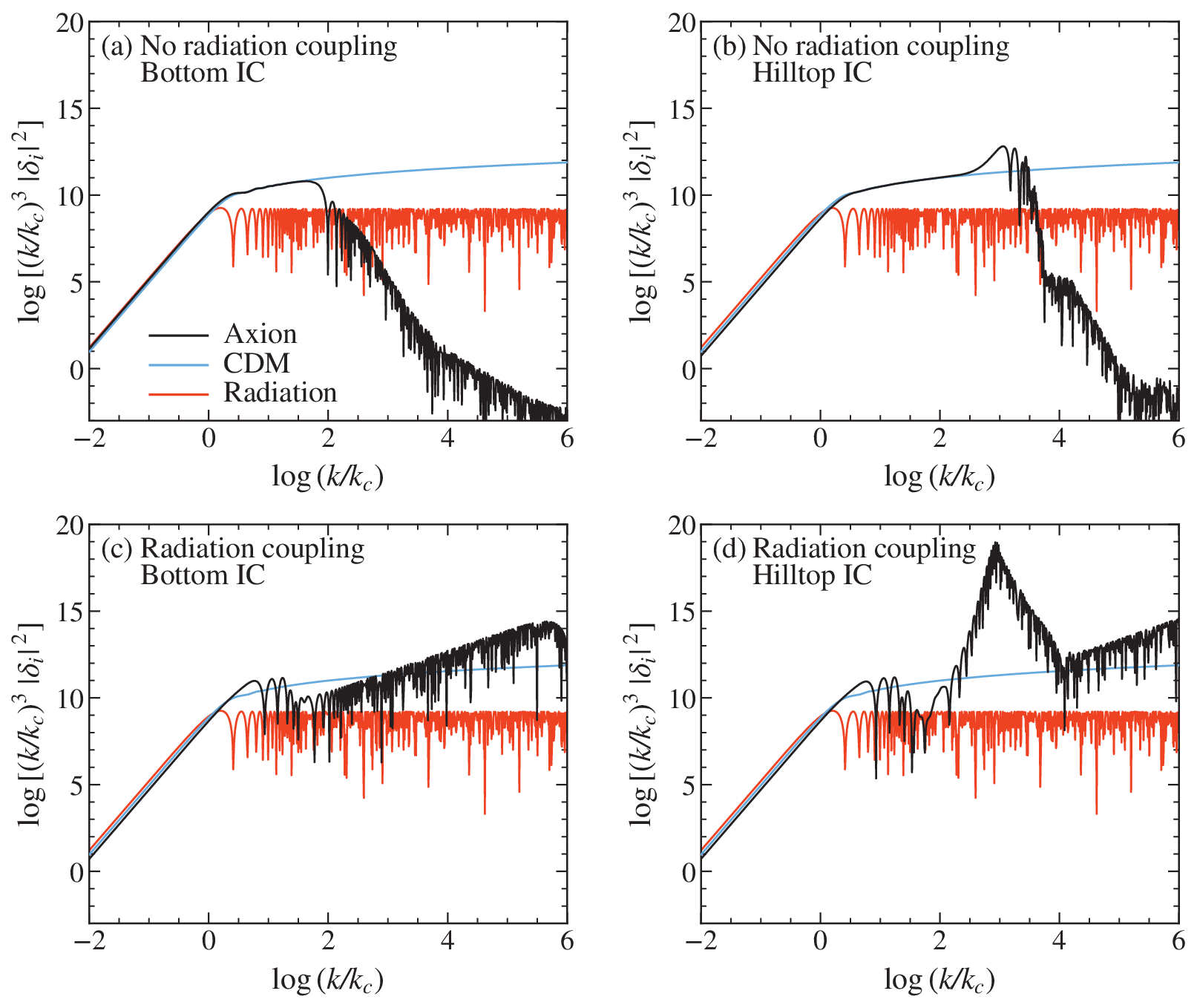}
\caption{
The axion-density power spectrum evaluated at $a_{\rm end}$. (a) $q=0$ and $\phi_i/f=0.1 \pi$. (b) $q=0$ and $\phi_i/f = (1 - 10^{-5}) \pi$. (c) $q=2.05$ and $\phi_i/f=0.1 \pi$. (d) $q=2.05$ and $\phi_i/f = (1 - 10^{-5}) \pi$. Black, blue, and red lines show the axion, CDM, and radiation density power spectra, respectively; $k/k_c \equiv kc/(a_c H_c) = (1.498 \times 10^5 {\rm AU}) (k/a_0)$, thus $k/k_c \sim 1$ corresponds to $k \sim (0.73\; {\rm parsec})^{-1}$ in the comoving size.
            }
\label{PS}
\end{figure}

The axion-density perturbation $\delta_\phi \equiv \delta \varrho_\phi/\varrho_\phi$ can be constructed using
\bea
   \delta \varrho_\phi
       = \dot \phi \delta \dot \phi
       + V_{,\phi} \delta \phi 
       - \dot \phi^2 \alpha
       + V_{,\varrho_r} \delta \varrho_r,
   \label{delta_phi-def}
\eea
which follows from Eq.\ (\ref{fluid-MSF-pert}). This represents the axion's contribution to the energy density in the energy-momentum tensor. Notice the presence of the $\delta \varrho_r$ contribution due to the direct coupling between radiation and the axion field in the potential. In radiation comoving gauge, from Eq.\ (\ref{eq7-r}) we have $\alpha = - {1 \over 4} \delta_r$.

Similarly to the background case, when the background and perturbed axion fields oscillate too rapidly to be handled numerically, we use the WKB approximation. For $\phi/f \ll 1$, Eq.\ (\ref{EOM-pert-RCG}) gives
\bea
   & & \delta \ddot \phi + 3 H \delta \dot \phi
       + \omega^2 \delta \phi = H^2 S, \quad
       \omega \equiv \sqrt{m^2 + {k^2 \over a^2}},
   \nonumber \\
   & & H^2 S = {1 \over 2} \dot \phi \dot \delta_r
       + \Big( {1 \over 2} 
       - {X_{,\varrho_r} \over X} \varrho_r \Big) m^2 \phi \delta _r.
\eea
For $H^2/\omega^2 \ll 1$, the WKB solution of the homogeneous part is \cite{Sikivie-Xue-2022}
\bea
   \delta \phi_{\rm H} = {A \over \sqrt{\omega} a^{3/2}} 
        \cos{(\Omega + B )}, \quad
        \Omega \equiv \int^t_{t_{\rm tr}} \omega d t.
\eea
The general solution is $\delta \phi = \delta \phi_{\rm H} + \delta \phi_{\rm IH}$ with the inhomogeneous part given as
\bea
   & & \delta \phi_{\rm IH}
        = \delta \phi_1 \int_{t_{\rm tr}}^t S H^2 a^3 \delta \phi_2 dt
        - \delta \phi_2 \int_{t_{\rm tr}}^t S H^2 a^3 \delta \phi_1 dt,
   \nonumber \\
   & &
   \delta \phi_1 \equiv {\sin{\Omega} \over \sqrt{\omega} a^{3/2}}, 
         \quad
         \delta \phi_2 \equiv {\cos{\Omega} \over \sqrt{\omega} a^{3/2}}.
   \label{IH-WKB}
\eea 
The constant coefficients $A$ and $B$ are determined by matching $\delta \phi$ and $\delta \dot \phi$ at the transition epoch.

\begin{figure}[t]
\hspace{-0.4cm} \vspace{.0cm}
\includegraphics[width=85mm,angle=0]{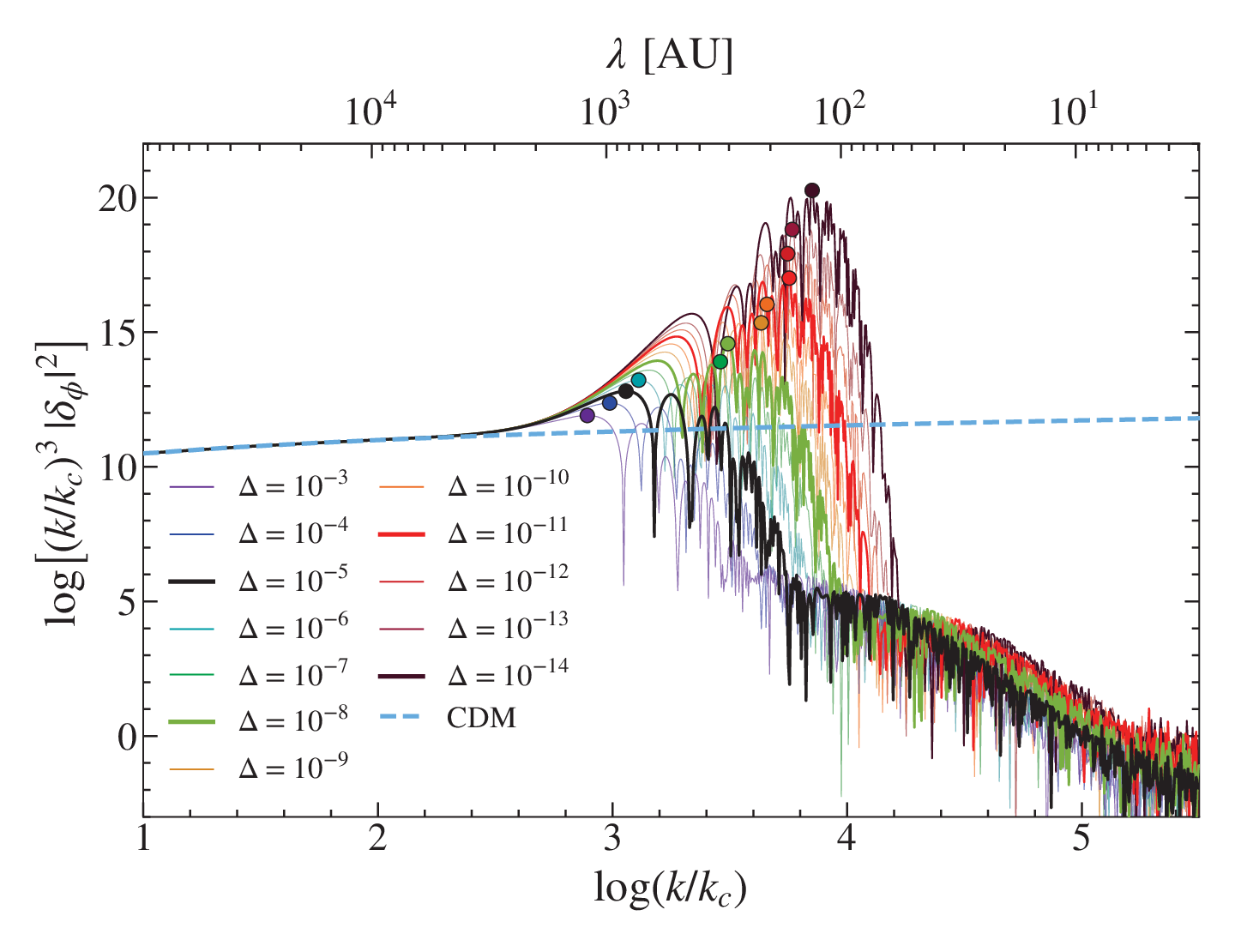}
\vspace{-0.0cm}
\caption{Enhancement in the case of Fig.\ \ref{PS}(b) with $q = 0$ and hilltop ICs; $\phi_i/f=(1-\Delta)\pi$. Peak positions are marked by dots. Compared with the CDM, at $\Delta = 10^{-14}$ the peak position reaches $|\delta_\phi|/|\delta_{\rm CDM}| \sim 2.4 \times 10^4$ accompanying a steep cliff in the small-scale side. The comoving wavelength $\lambda \equiv 2 \pi a_0/k$.
            }
\label{P-q=0}
\end{figure}

For $\phi/f \ll 1$, $H/m \ll1$ and $k^2/a^2 \ll m^2$, and further ignoring the oscillatory part by time average, we have
\bea
   & & \delta_\phi = 2 {A \over \overline A} 
       \cos{(\Omega - \overline \Omega + B - \overline B)}
   \nonumber \\
   & & \qquad
      + {1 \over 4} \Big(
      3 + 2 {X_{,\varrho_r} \over X} \varrho_r \Big)\delta_r
      - {1 \over 2} \delta_r (t_{\rm tr}).
   \label{delta_phi-WKB-sol}
\eea
Asymptotically we recover $\delta_\phi = {3 \over 4} \delta_r$ as expected in the CDM case. For more accurate forms of the WKB approximation used in the numerical study, without assuming $\omega = m$ and including higher-order terms in $H/m$, see Eqs.\ (\ref{delta-phi-WKB1})-(\ref{delta-phi-WKB4}).

For time-dependent $g_*$ and $g_{*s}$, $p_r = {1 \over 3} \varrho_r$ is no longer valid, and we have to solve the radiation and axion perturbation equations in a more general case. The perturbation equation of a general ideal fluid in comoving gauge is derived in \cite{Bardeen-1980, Hwang-1999}
\bea
   & & \hskip -1cm 
       \ddot \delta_r + ( 2 + 3 c_s^2 - 6 w) H \dot \delta_r
   \nonumber \\
   & & \hskip -1cm \qquad
       - \Big[ 4 \pi G \varrho ( 1 - 6 c_s^2 + 8 w - 3 w^2 )
       + c_s^2 {\Delta \over a^2} \Big] \delta_r = 0,
   \label{ddot-delta-eq-w}
\eea
where $w$ and $c_s^2$ for the radiation plasma are given in Eq.\ (\ref{w-c_s}), and we define $\delta p_r \equiv c_s^2 \delta \varrho_r + e$ with $e \equiv 0$, which corresponds to neglecting the entropic perturbation of the radiation. The equation of motion is given by
\bea
   & & \hskip -1cm
       \delta \ddot \phi + 3 H \delta \dot \phi
       - {\Delta \over a^2} \delta \phi
       + V_{,\phi\phi} \delta \phi
   \nonumber \\
   & & \hskip -1cm \qquad
       = \dot \phi ( \kappa + \dot \alpha )
       - ( 2 V_{,\phi} + 3 H \dot \phi ) \alpha
       - V_{,\phi\varrho_r} \varrho_r \delta _r \equiv H^2 S,
\eea
with 
\bea
   \alpha = - {c_s^2 \over 1+ w} \delta_r, \quad 
       \kappa = {1 \over 1+w}( \dot \delta_r - 3 w H \delta_r),
   \label{alpha-kappa}
\eea
which follow from Eqs.\ (\ref{eq6})-(\ref{EOM-pert}). For $p_r = {1 \over 3} \varrho_r$, we have $w = {1 \over 3} = c_s^2$ and recover Eqs.\ (\ref{ddot-delta-RCG})-(\ref{EOM-pert-RCG}).

As the IC we impose a near Harrison-Zel'dovich spectrum \cite{Harrison-1970, Zeldovich-1972} which is also motivated by inflationary theory \cite{Mukhanov-Chibisov-1981} and observations of the cosmic background radiation \cite{Planck-2020}. At $a_i = 10^{-18} a_0$ we use $\delta_{r} = {4 \over 3} C x^2$; for ICs of $\phi$ and $\delta \phi$, see Eq.\ (\ref{ICs}). The near Harrison-Zel'dovich spectrum corresponds to $|C|^2 \propto k^{n_S - 4}$ with $n_S \simeq 1$. Since we compare the CDM and radiation power spectra simultaneously, we do not impose an observational normalization on the axion density perturbation; for equations of CDM in radiation era, see Eq.\ (\ref{delta-eq-CDM}).

\begin{figure}[t]
\hspace{-0.4cm} \vspace{.0cm}
\includegraphics[width=85mm,angle=0]{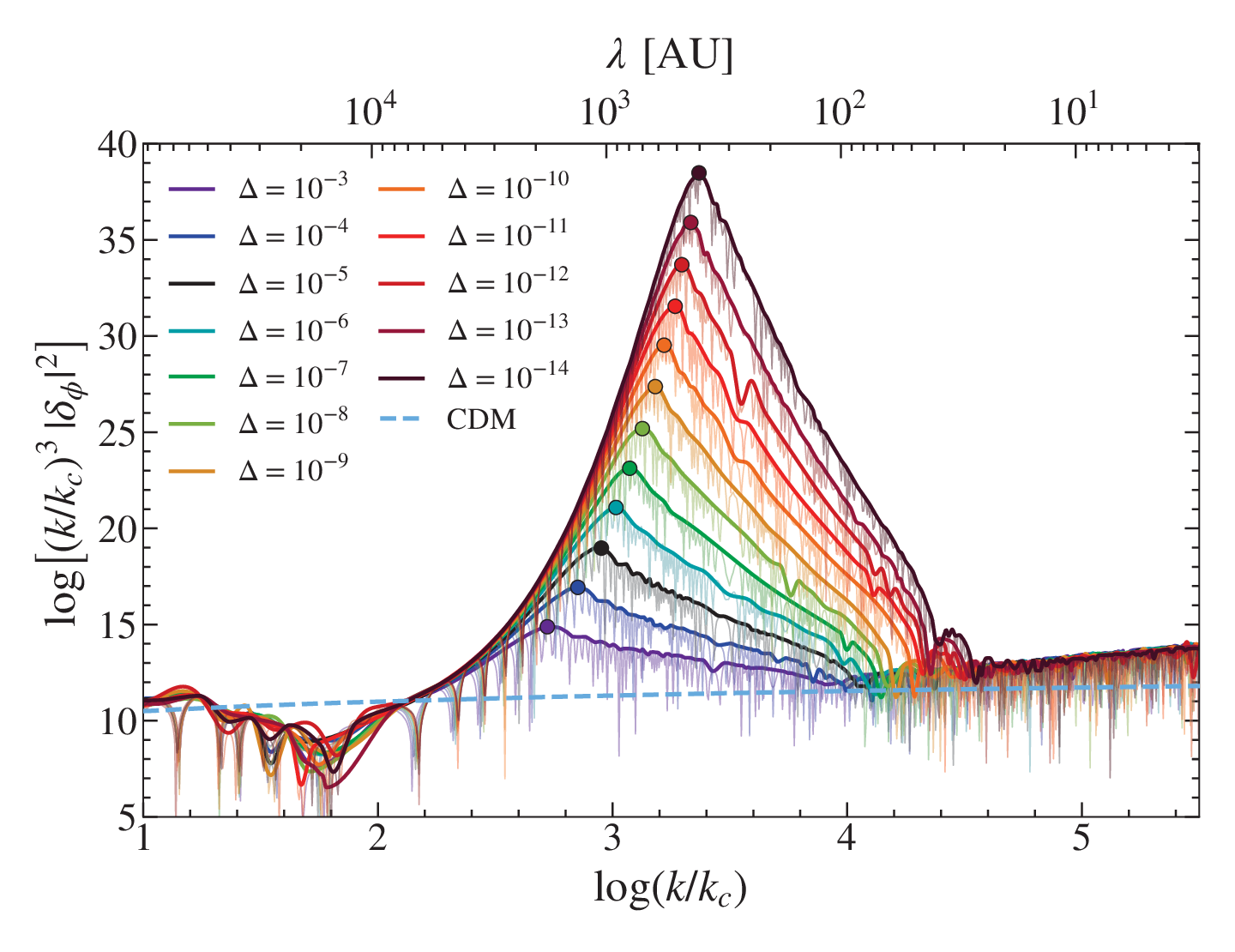}
\vspace{-0.0cm}
\caption{Axion pyramids in the case of Fig.\ \ref{PS}(d) with $q= 2.05$ and hilltop ICs. Envelopes and peak positions are added. Quantitative relations are given in Table \ref{table} and Fig.\ \ref{P2}. 
            }
\label{P}
\hspace{-0.4cm} \vspace{.0cm}
\includegraphics[width=44mm,angle=0]{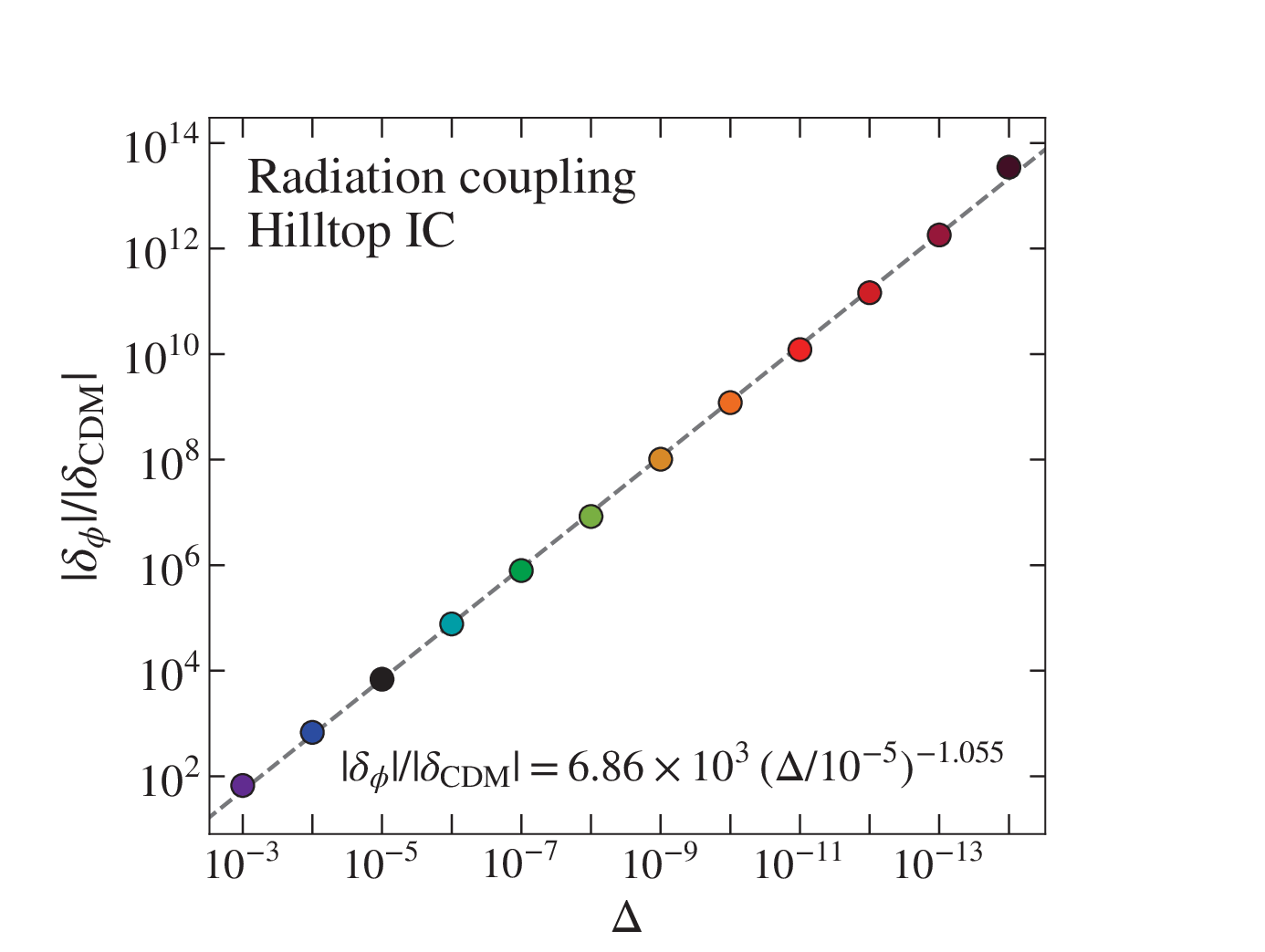}
\includegraphics[width=44mm,angle=0]{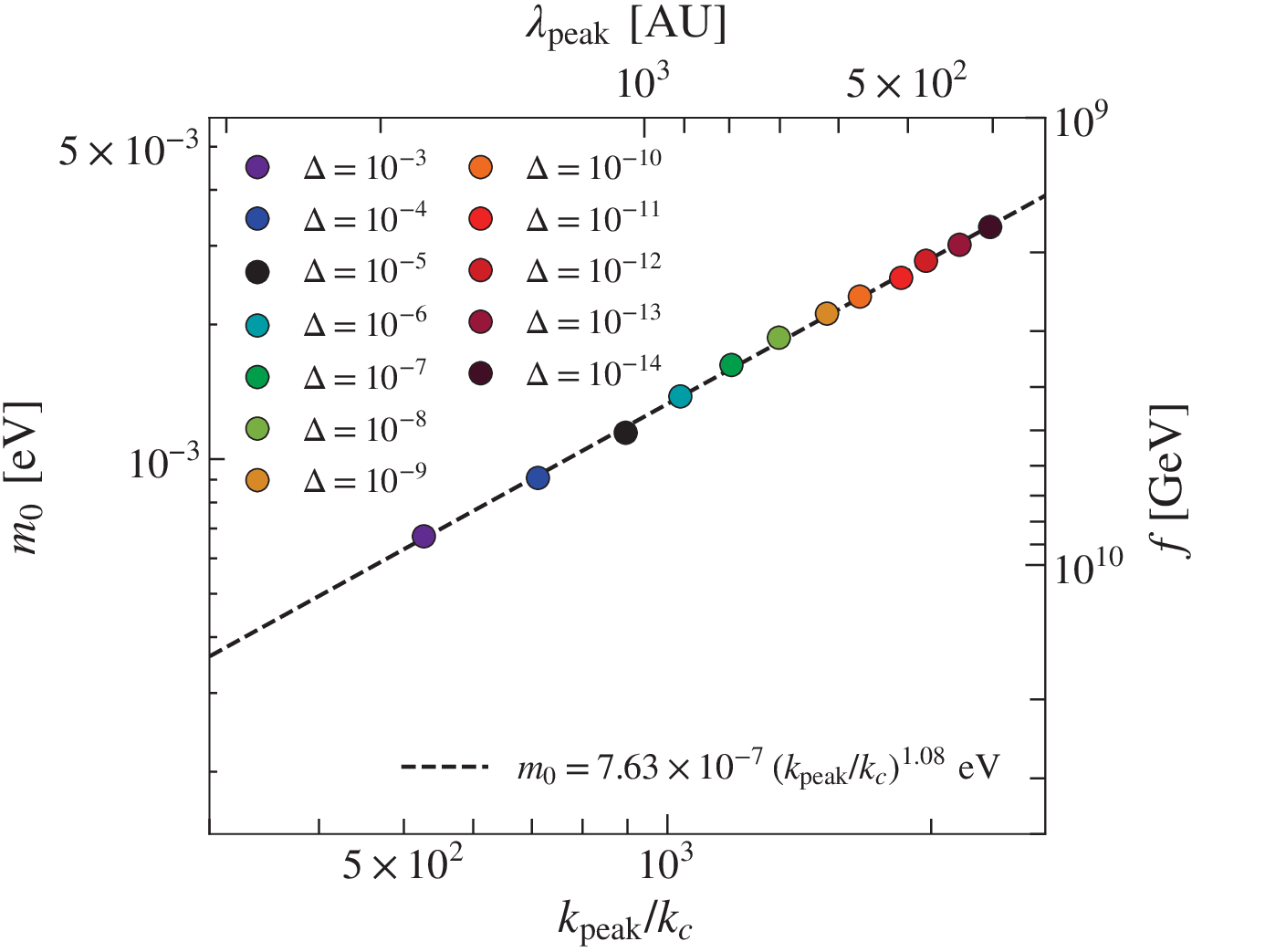}
\caption{Left: Peak amplitude of the axion density perturbation
as a function of the hilltop IC. Right: Peak location in $k$-space as a function of the axion mass. Observation of the axion pyramid can constrain the axion mass.
            }
\label{P2}
\end{figure}
\begin{table}[t!]
\centering
\caption{Peak location and height (relative to CDM) of the axion pyramids in the radiation coupled QCD axion with hilltop ICs. The peak locations are determined by fitting a smooth spline to the upper envelope of the oscillating spectrum in Fig.\ \ref{P}.}
\label{table}
\begin{tabular}{cccc}
\hline
$\Delta$ & \qquad $k_{\rm peak}/k_c$ \qquad & \quad $\lambda_{\rm peak}[{\rm AU}]$ & \qquad $|\delta_\phi|/|\delta_{\rm CDM}|$ \qquad \\
\hline
  $10^{-3}$ & \quad 527 & 1785 & \qquad $6.68\times10^{1}$ \\
  $10^{-5}$ & \quad 896 & 1050 & \qquad $6.88\times10^{3}$ \\
  $10^{-8}$ & \quad 1340 & 702 & \qquad $8.35\times10^{6}$ \\
  $10^{-11}$ & \; \, 1848  & 509 & \qquad\, $1.22\times10^{10}$  \\
  $10^{-14}$ & \; \, 2335  & 403 & \qquad\, $3.48\times10^{13}$  \\
\hline
\end{tabular}
\end{table}

\section{Results}

We present the axion density power spectrum, ${\cal P}_{\delta_\phi} \equiv {k^3 \over 2 \pi^2} |\delta_\phi|^2$, in four cases characterized by two parameters, $q$ (radiation coupling) and $\phi_i/f$ (self-coupling).
(a) $q = 0$ and bottom IC: an ordinary massive axion with constant mass.
(b) $q = 0$ and hilltop IC: an axion with sine-Gordon potential and constant mass.
(c) $q = 2.05$ and bottom IC.
(d) $q = 2.05$ and hilltop IC.
The hilltop IC activates the effect of self-coupling, and nonvanishing $q$ causes a direct coupling between the axion and radiation.

The axion density power spectra for the four cases are presented in 
Fig.\ \ref{PS}. The parameter $f$ (thus, $m_0$) is determined for a given initial field position $\phi_i/f$ by matching the axion background density after the phase transition to the CDM density: $\varrho_\phi a_{\rm end}^3 = \varrho_{\rm CDM} a_{\rm end}^3$. In this work, we set the endpoint to
$a_{\rm end} = 4.712 \times 10^{-12} a_0$ where $X = 0.99988$ ($T_\textrm{end}=49.83~\textrm{MeV}$), after which $\varrho_\phi a^3 = \varrho_{\phi 0}a_0^3 = (3H_0^2/8\pi G)\Omega_\phi a_0^3$. We adopt $\Omega_\phi=\Omega_{\rm CDM}=0.26$ and $h=0.677$, corresponding to $\Omega_\phi h^2=0.12$ \cite{Planck-2020}, where $H_0\equiv100h\,{\rm km\,s^{-1}\,Mpc^{-1}}$.

Figure \ref{PS}(a) shows suppression of the power spectrum for a massive axion without self-coupling. It is a well-studied case of the axion as CDM in the background \cite{Preskill-1983, Abbott-1983, Dine-1983}, with non-relativistic perturbations \cite{Khlopov-1985, Bianchi-1990}, and with relativistic perturbations \cite{Sikivie-Yang-2009, Hwang-2009, Noh-2017}. The current quantum Jeans scale caused by the presence of quantum stress is \cite{Hu-2000, Sikivie-Yang-2009, Hwang-2009}
\bea
   & & \lambda_J = {2 \pi a_0 \over k_J}
       = \sqrt{{\pi \hbar \over m} \sqrt{\pi \over G \varrho_{\phi 0}}}
   \nonumber \\
   & & \qquad
       = 96.4 \Big( {f \over 10^{12} {\rm GeV}} \Big)^{1/2}
       \Big( {\Omega_{\phi}h^2 \over 0.12} \Big)^{-1/4} {\rm AU}.
\eea
In the relativistic calculation, the above quantum Jeans scale is valid in axion-comoving gauge, zero-shear gauge and uniform-curvature gauge \cite{Sikivie-Yang-2009, Hwang-2009, Hwang-2022}. The small-scale suppression in the radiation-dominated QCD phase transition era is numerically noticed in Fig.\ 5 of \cite{Arvanitaki-2020}.

\begin{figure}[t]
\hspace{-0.2cm} \vspace{.0cm}
\includegraphics[width=83mm,angle=0]{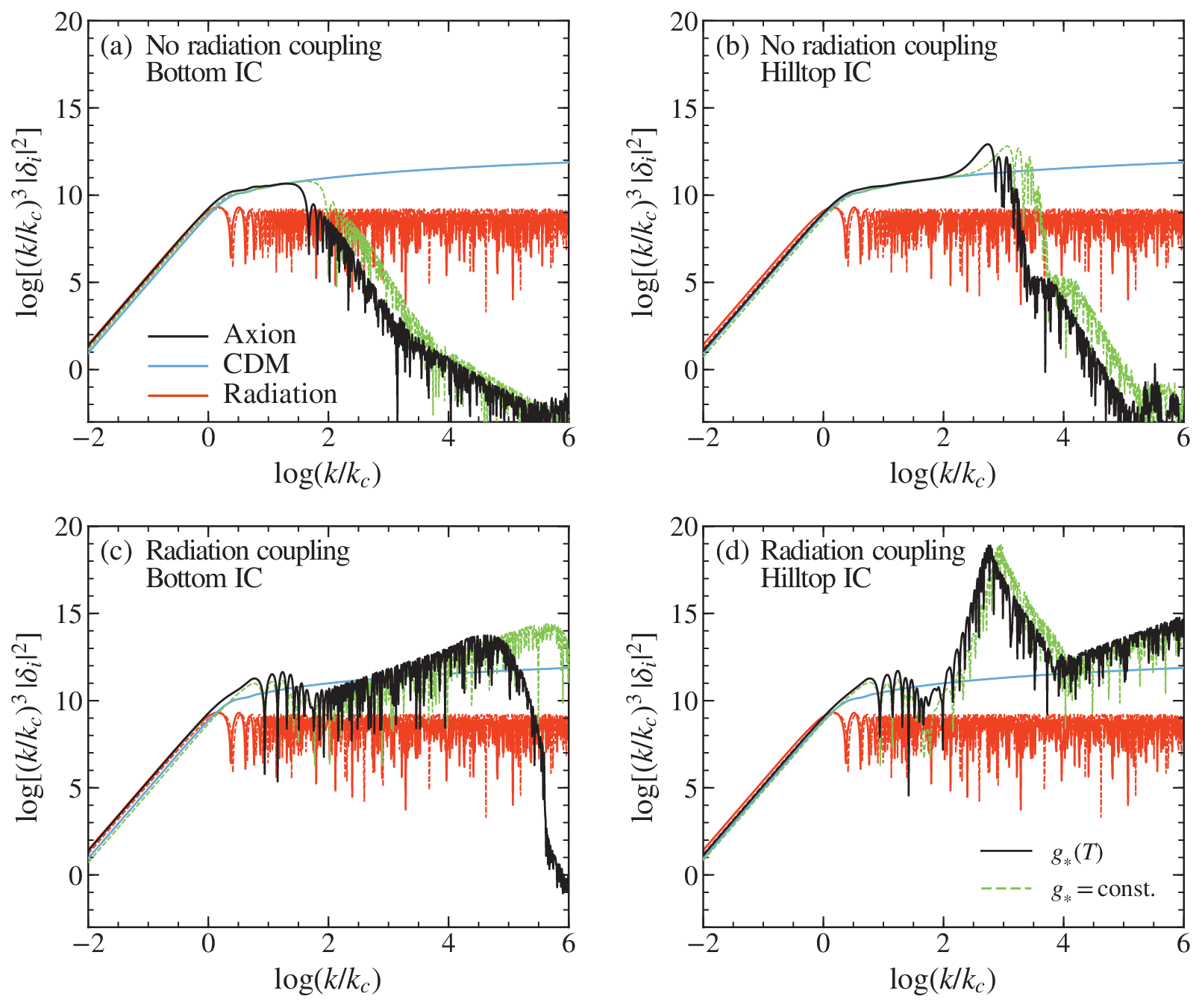}
\caption{Same as in Fig.\ \ref{PS}, but for time-dependent $g_*$ and $g_{*s}$. The cases with constant $g_*$ are also presented in green for axions and with dashed lines for photons and CDM; $k/k_c \equiv kc/(a_c H_c) = (0.985 \times 10^5 {\rm AU}) (k/a_0)$, thus $k/k_c \sim 1$ corresponds to $k \sim (0.48\; {\rm parsec})^{-1}$ in terms of the comoving length scale. The normalization of $k_c$ for time-dependent $g_*$ is larger by a factor of $1.52$ compared with that for constant $g_*$, thereby causing substantial shifts in the axion spectra on small scales; the differences in physical scales, however, are much smaller, as can be seen by comparing Fig.\ \ref{P} and Table \ref{table} with Fig.\ \ref{P-g} and Table \ref{table-g}.
            }
\label{PS-g}
\end{figure}

Figure \ref{PS}(b) is a case with $q = 0$ and the hilltop IC. The small-scale suppression persists, with a slight enhancement near the transition scale relative to the CDM case. The enhancement becomes larger as the IC approaches the hilltop, see Fig.\ \ref{P-q=0}. This case was studied in Figs.\ 4 and 5 of \cite{Arvanitaki-2020} in the zero-shear (conformal-Newtonian) gauge setting $\chi \equiv 0$. Figure \ref{PS}(c) is a case with $q=2.05$ and the bottom IC. Compared with the CDM case, the axion power spectrum shows a blue slope. 

Figure \ref{PS}(d) reveals a conspicuous and sharp enhancement of the power spectrum in the case of $q= 2.05$ and the hilltop IC. This case was studied in \cite{Arvanitaki-2020, Kitajima-2022} in different gauge conditions; see Fig.\ 17 of \cite{Arvanitaki-2020} in zero-shear gauge, and Figs.\ 10 and 11 of \cite{Kitajima-2022} in the uniform-radiation gauge setting $\delta \varrho_r \equiv 0$. However, in all previous numerical studies the evolution was terminated before fully covering the effect of the QCD phase transition which necessitates the WKB procedure.

\begin{figure}[t]
\hspace{-0.4cm} \vspace{.0cm}
\includegraphics[width=85mm,angle=0]{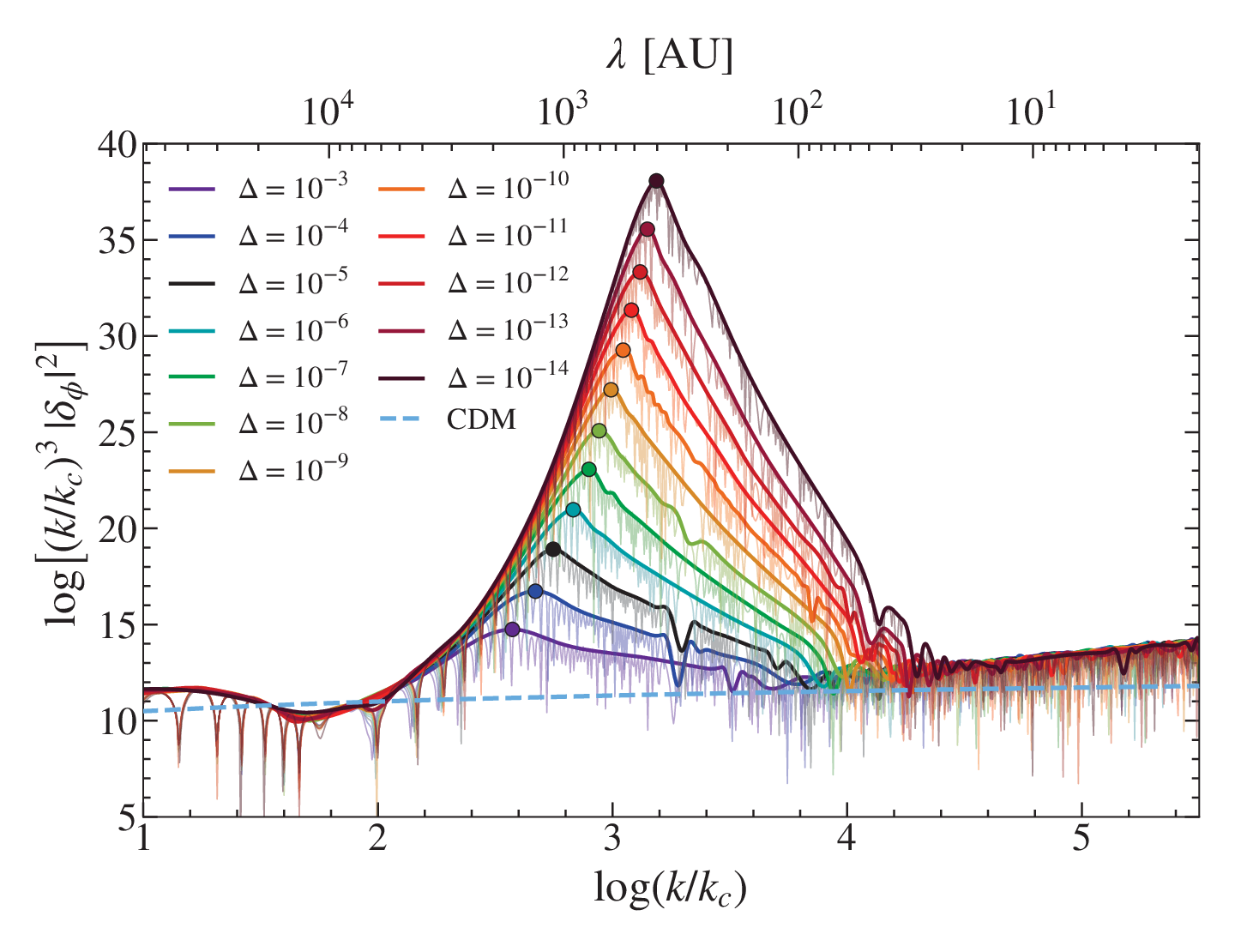}
\vspace{-0.0cm}
\caption{Same as Fig.\ \ref{P}, but for time-dependent $g_*$ and $g_{*s}$.
            }
\label{P-g}
\hspace{-0.4cm} \vspace{.0cm}
\includegraphics[width=44mm,angle=0]{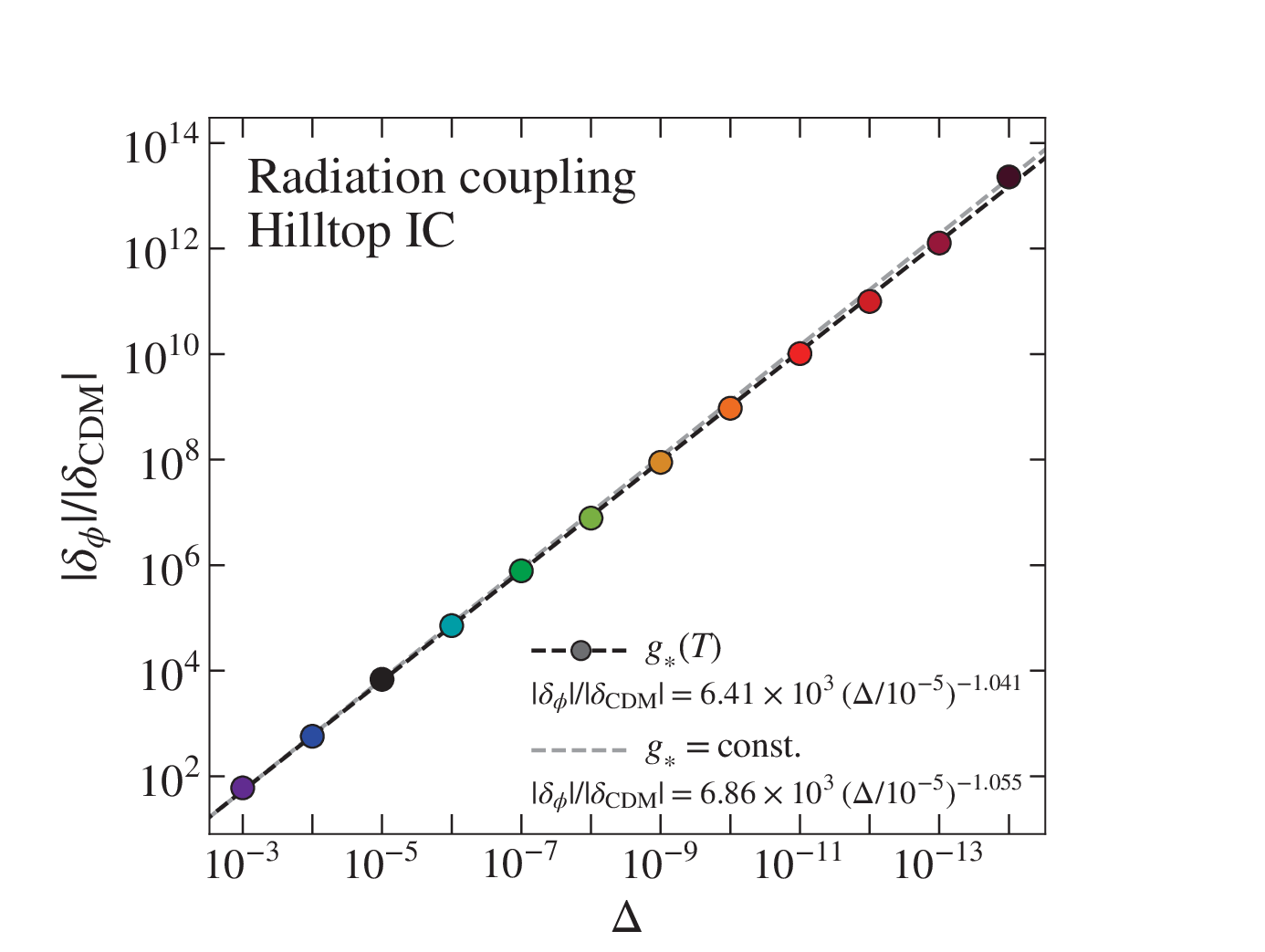}
\includegraphics[width=44mm,angle=0]{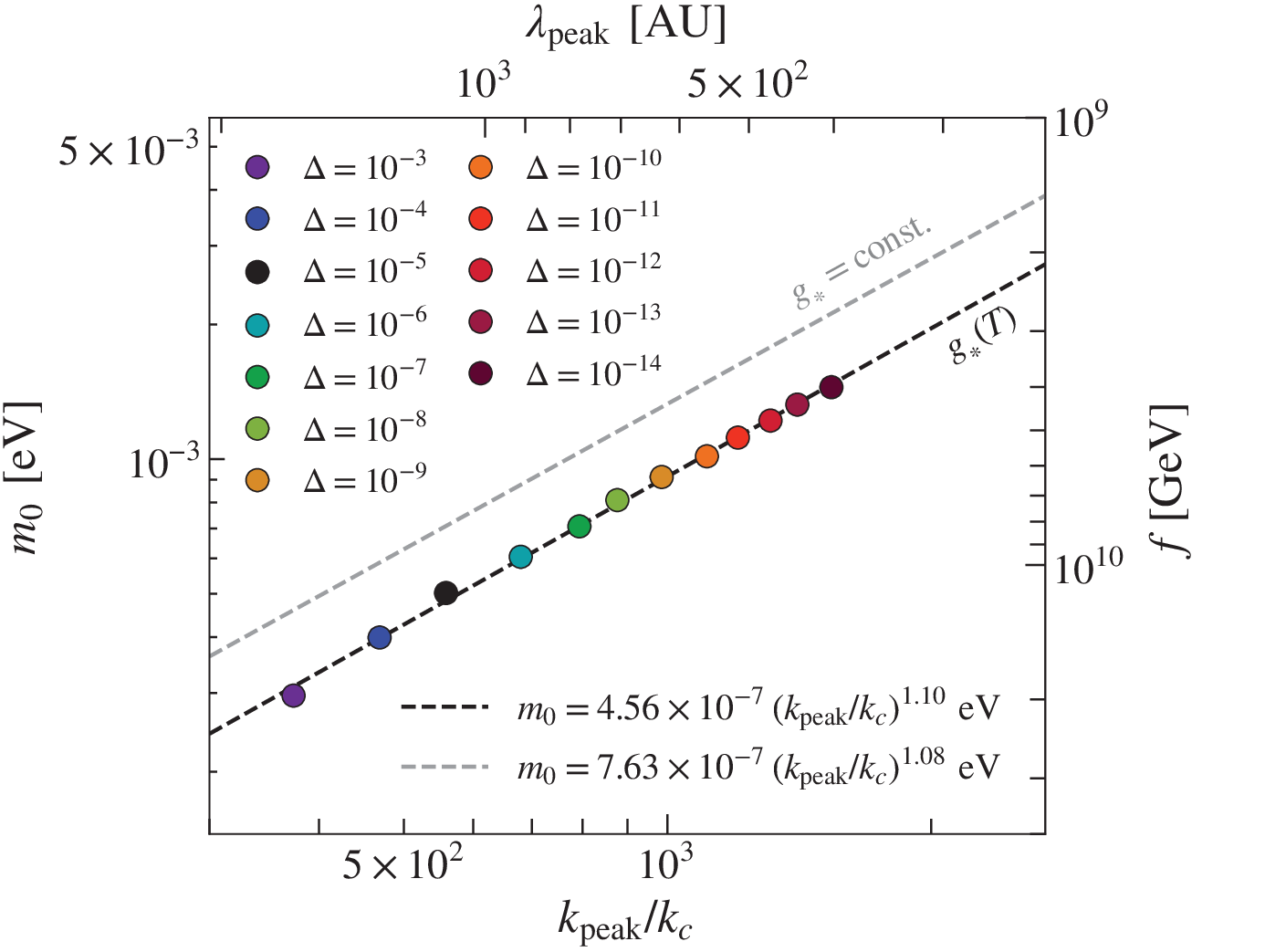}
\caption{Same as Fig.\ \ref{P2}, but for time-dependent $g_*$ and $g_{*s}$. The $f$ versus $\Delta$ (and thus $\phi_i/f$) relation in this case approximately follows the one in Fig.\ 16 of \cite{Arvanitaki-2020}. 
            }
\label{P2-g}
\end{figure}

\begin{table}[b]
\centering
\caption{Same as Table \ref{table}, but for time-dependent $g_*$ and $g_{*s}$ in Fig.\ \ref{P-g}.}
\label{table-g}
\begin{tabular}{lccc}
\hline
\; $\Delta$ & \qquad $k_{\rm peak}/k_c$ \qquad & \quad $\lambda_{\rm peak}[{\rm AU}]$ & \qquad $|\delta_\phi|/|\delta_{\rm CDM}|$ \qquad \\
\hline
  $10^{-3}$ & \quad 374 & 1655 & \qquad $6.00\times10^{1}$ \\
  $10^{-5}$ & \quad 559 & 1108 & \qquad $6.86\times10^{3}$ \\
  $10^{-8}$ & \quad 877 & 706 & \qquad $7.78\times10^{6}$  \\
 $10^{-11}$ & \quad 1203 & 515 & \qquad\, $1.02\times10^{10}$ \\
  $10^{-14}$ & \quad 1539 & 402 & \quad\;\;\;\,\, $2.27\times10^{13}$ \\
\hline
\end{tabular}
\end{table}

The peak in Fig.\ \ref{PS}(d) appears around $k_{\rm peak}/k_c \sim 900$, thus $\lambda_{\rm peak}\equiv2 \pi a_0/k_{\rm peak} \sim 1050{\rm AU}$ with a peak enhancement relative to CDM of $|\delta_\phi|/|\delta_{\rm CDM}| \sim 6.9 \times 10^3$. We may call it an {\it axion pyramid}. The enhancement increases as the IC is set closer to the hilltop \cite{Arvanitaki-2020, Kitajima-2022}; see Table 1 and Figs.\ \ref{P} and \ref{P2}. With $\phi_i/f=(1-\Delta)\pi$, our code can reach up to $\Delta = 10^{-14}$, yielding a $10^{27}$-fold amplification in the power spectrum, thus $|\delta_\phi|/|\delta_{\rm CDM}| \sim 3.5 \times 10^{13}$. By fitting the peak positions we obtain a relation $|\delta_\phi|/|\delta_{\rm CDM}| \sim 6.86 \times 10^3 (\Delta/10^{-5})^{-1.055}$.

In Figs.\ \ref{PS}-\ref{P2}, we take $p_r = {1 \over 3} \varrho_r$ with $g_* = 3.38$ accounting only for photons and massless neutrino species. By properly taking into account the time dependence of $g_*$ and $g_{*s}$, we no longer have $p_r = {1\over 3} \varrho_r$, and we have to solve Eqs.\ (\ref{ddot-delta-eq-w})-(\ref{alpha-kappa}) with Eq.\ (\ref{w-c_s}). In this case, for the background evolution, the axion abundance curve (the $f$ versus $\phi_i/f$ relation) in Fig.\ 16 of \cite{Arvanitaki-2020} is recovered; however, in calculating the power spectra, \cite{Arvanitaki-2020, Kitajima-2022} adopted $p_r = {1 \over 3} \varrho_r$.

In Figs.\ \ref{PS-g}-\ref{P2-g} and Table \ref{table-g}, we present the cases shown in Figs.\ \ref{PS}-\ref{P2} and Table \ref{table}, with time-dependent $g_*$ and $g_{*s}$. With the same end temperature $T_\textrm{end}$ ($X = 0.99997$ for $q=2.05$), we have $a_{\rm end} = 3.064\times 10^{-12} a_0$, and $k_c$ is also shifted. For the cases in Figs.\ \ref{PS}(a)-(b) with $q = 0$, however, when $g_*$ and $g_{*s}$ are time-dependent, the power spectra show numerical errors in the large-$k$ range. For constant $g_*$, the small-scale damping due to the quantum stress often involves a cancellation between the first two terms in Eq.\ (\ref{delta_phi-def}), as well as a further cancellation involving the third term; such delicate cancellations among rapidly oscillating terms are not easily maintained for time-dependent $g_*$ and $g_{*s}$. For the hilltop IC with $q = 2.05$, the results in Figs.\ \ref{P-g}, \ref{P2-g} and Table \ref{table-g} show only minor quantitative changes in the position and height of the pyramid between the cases with time-dependent and constant $g_*$.

\section{Discussion}

Some enhancements during the QCD phase transition were previously interpreted as being caused by parametric resonance \cite{Arvanitaki-2020, Kitajima-2022}. 
In the oscillating regime, the left-hand side of Eq.\ (\ref{EOM-pert-RCG}) can be arranged into a damped Mathieu equation \cite{Greene-1999}. However, the enhancement in Fig.\ \ref{PS}(d) appears in $\delta_\phi$, which is a combination of background and perturbed quantities defined in Eq.\ (\ref{delta_phi-def}), whereas $\delta\phi$ and $\delta\dot\phi$ exhibit less dramatic changes during the enhancement. In the numerical evolution, the pyramid is already established before the WKB transition, which is imposed in the rapid oscillation regime but occurs {\it prior} to the QCD phase transition during which substantial axion mass generation occurs. This may explain why the enhancement was previously observed in \cite{Arvanitaki-2020, Kitajima-2022}. A proper theoretical understanding of the formation of the axion pyramid remains an interesting subject for further study.

In this work, we showed that the QCD phase transition during the radiation-dominated era can cause a significant enhancement in the axion density power spectrum depending on the radiation coupling in the axion mass and the IC in the axion potential. We adopted the radiation-dominated approximation, which is sufficient during the QCD phase transition era. More importantly, we adopted the radiation-comoving gauge as the temporal gauge (slicing or hypersurface) condition. The axion density power spectrum {\it depends} on the temporal gauge choice \cite{Hwang-2022}. Results in other gauges will be studied in future work. Here, we note that direct numerical calculations, although unable to cover the entire process, in other gauge conditions still exhibit and preserve the axion-pyramid feature.

Diverse enhancement mechanisms of axion DM structures during the QCD phase transition were suggested previously, and there are studies concerning the phenomenology of axion minicluster/clump and their potential observational consequences, see \cite{Hogan-1988, Kolb-1996, Eroncil-2023, Arvanitaki-2020}. Here, we have quantified the changes in the power spectrum caused by the QCD phase transition. The implications of the axion pyramid for subsequent astrophysical and cosmological phenomenology and its potential observational consequences are left for future investigation.

\section*{Acknowledgments}

We thank Professors Kiwoon Choi and Wan Il Park and Drs.\ Gongjun Choi, Heejung Kim and Hyeonseok Seong for helpful discussions. J.H.\ was supported by IBS under the project code, IBS-R018-D1. H.N.\ was supported by the National Research Foundation (NRF) of Korea funded by the Korean Government (No.\ RS-2024-00333721). C.-G.P. was supported by research funds of Jeonbuk National University in 2025 and the National Research Foundation of Korea (NRF) grant funded by the Korea government (MSIT) No. RS-2026-25473390.

%
%


\begin{widetext}
\appendix
%
%
%
\section{Perturbation equations}
                                      \label{Appendix-eqs}

Our metric convention is in Eq.\ (\ref{metric}). We consider a system with radiation and a scalar field in a flat Friedmann background. We include a direct coupling between the radiation and scalar field with $V = V(\phi, \varrho_r)$. To the background order, we have Eq.\ (\ref{fluid-BG}).

To the perturbed order, without imposing a temporal gauge condition, the complete set of equations is as follows. Conservation equations for the radiation give
\bea
   & & ( 1 + V_{,\varrho_r} ) \delta \dot \varrho_r
       + 4 H \delta \varrho_r
       = {4 \over 3} \varrho_r
       \Big( \kappa - 3 H \alpha
       + {\Delta \over a} v_r \Big)
       - \dot \varrho_r ( V_{,\phi\varrho_r} \delta \phi
       + V_{,\varrho_r\varrho_r} \delta \varrho_r ),
   \label{eq6-r} \\
   & & {1 \over a^{4}} \left( a^4 \varrho_r v_r
       \right)^{\displaystyle\cdot}
       = {1 \over a} \Big( {1 \over 4} \delta \varrho_r
       + \varrho_r \alpha
       - {3 \over 4} V_{,\varrho_r} \delta \varrho_r \Big),
    \label{eq7-r}
\eea
where $\kappa$ is the perturbation of the trace of extrinsic curvature in Eq.\ (\ref{eq1}); we ignore anisotropic stress of the radiation. For a single-component fluid without anisotropic stress, the conservation equations are
\bea
   & & \delta \dot \varrho
       + 3 H ( \delta \varrho + \delta p )
       = ( \varrho + p )
       \Big( \kappa - 3 H \alpha
       + {\Delta \over a} v \Big),
   \label{eq6} \\
   & & {1 \over a^{4}} [ a^4 ( \varrho + p ) v ]^{\displaystyle{\cdot}}
       = {1 \over a} ( \varrho + p ) \alpha
       + {1 \over a} \delta p.
    \label{eq7}
\eea
The perturbed scalar field equation of motion gives
\bea
   \delta \ddot \phi + 3 H \delta \dot \phi
       - {\Delta \over a^2} \delta \phi
       + V_{,\phi\phi} \delta \phi
       + V_{,\phi\varrho_r} \delta \varrho_r
       = \dot \phi \left( \kappa + \dot \alpha \right)
       + ( 2 \ddot \phi + 3 H \dot \phi ) \alpha.
   \label{EOM-pert}
\eea
The metric parts are described by Einstein's equations \cite{Bardeen-1988}
\bea
   & & \kappa \equiv 3 H \alpha - 3 \dot \varphi
       - {\Delta \over a^{2}} \chi,
   \label{eq1} \\
   & & H \kappa
       + {\Delta \over a^{2}} \varphi
       = - 4 \pi G \delta \varrho,
   \label{eq2} \\
   & & \kappa + {\Delta \over a^{2}} \chi
       = 12 \pi G \left(
       \varrho + p \right) a v,
   \label{eq3} \\
   & & \dot \kappa + 2 H \kappa
       + \Big( 3 \dot H + {\Delta \over a^{2}} \Big)
       \alpha = 4 \pi G \left( \delta \varrho
       + 3 \delta p \right),
   \label{eq4} \\
   & & \varphi + \alpha
       - \left( \dot \chi + H \chi \right)
       = 0,
   \label{eq5}
\eea
where the collective fluid quantities are given in terms of individual ones as
\bea
   & & \delta \varrho = \delta \varrho_r
       + \delta \varrho_\phi, \quad
       \delta p = {1 \over 3} \delta \varrho_r
       + \delta p_\phi, \quad
       (\varrho + p) v = {4 \over 3} \varrho_r v_r
       + (\varrho_\phi + p_\phi) v_\phi,
   \label{fluid-pert}
\eea
with the fluid quantities of the scalar field
\bea
   & & \delta \varrho_\phi
       = \dot \phi \delta \dot \phi
       + V_{,\phi} \delta \phi
       - \dot \phi^2 \alpha
       + V_{,\varrho_r} \delta \varrho_r, \quad
       \delta p_\phi
       = \dot \phi \delta \dot \phi
       - V_{,\phi} \delta \phi
       - \dot \phi^2 \alpha
       - V_{,\varrho_r} \delta \varrho_r, \quad
       (\varrho_\phi + p_\phi) v_\phi
       = {1 \over a} \dot \phi \delta \phi.
   \label{fluid-MSF-pert}
\eea

The above set of equations is presented without imposing a temporal gauge (hypersurface or slicing) condition. Radiation comoving gauge imposes $v_r \equiv 0$ as the temporal gauge condition. Since this gauge condition completely removes the gauge degrees of freedom, all remaining perturbation variables are gauge-invariant.

\section{Equations in dimensionless forms}

\subsection{Background}
                                      \label{Appendix-BG}

We consider
\bea
   & & V = f^2 m_0^2 X [ 1 - \cos{(\phi/f)} ], \quad
       X (\varrho_r) = {1 \over 1 + (\varrho_r / \varrho_{rc})^q},
\eea
where $m_0 = 5.8 \times 10^{-6} {\rm eV} ( 10^{12} {\rm GeV}/f )$; for the QCD phase transition we have $q = 2.05$. Setting $c \equiv 1 \equiv \hbar$ and $8 \pi G \equiv 1$, $\phi$ and $f$ are in units of the reduced Planck energy with $E_{\rm Planck} \equiv \sqrt{\hbar c^5 \over 8 \pi G} = 2.436 \times 10^{18} {\rm GeV}$.
We have
\bea
   & & V_{,\phi} = f m_0^2 X \sin{(\phi/f)}, \quad
       V_{,\phi\phi} = m_0^2 X \cos{(\phi/f)}, \quad
       V_{,\phi \varrho_r} = V_{,\phi} X_{,\varrho_r}/X, \quad
       V_{,\varrho_r} = V X_{,\varrho_r}/X, \quad
       V_{,\varrho_r \varrho_r}
       = V X_{,\varrho_r \varrho_r}/X, 
   \nonumber \\
   & &
       X_{,\varrho_r}
       = - {q (\varrho_r/\varrho_{rc})^{q-1}/\varrho_{rc}
       \over [ 1 + (\varrho_r / \varrho_{rc})^q ]^2}, \quad
       X_{,\varrho_r \varrho_r}
       = {2 q^2 (\varrho_r/\varrho_{rc})^{2q-2}/\varrho_{rc}^2
       \over [ 1 + (\varrho_r / \varrho_{rc})^q ]^3}
       - {q (q-1) (\varrho_r/\varrho_{rc})^{q-2}/\varrho_{rc}^2
       \over [ 1 + (\varrho_r / \varrho_{rc})^q ]^2}.
   \label{V-X}
\eea

Using $\phi^\prime \equiv {d \phi \over d N}$ with $N \equiv \ln{(a/a_0)}$, background equations are
\bea
   & & \phi^{\prime\prime}
       + \left( 3 + {\dot H \over H^2} \right) \phi^\prime
       + {V_{,\phi} \over H^2} = 0, \quad
       \varrho_\phi = {1 \over 2} H^2 \phi^{\prime 2} + V, \quad
       p_\phi = {1 \over 2} H^2 \phi^{\prime 2} - V,
   \nonumber \\
   & &
       \varrho_\phi^\prime + 3 ( \varrho_\phi + p_\phi )
       = V_{,\varrho_r} \varrho_r^\prime, \quad
       \varrho_r^\prime = - {4 \over 1 + V_{,\varrho_r}} \varrho_r, 
   \nonumber \\
   & &
       {H^2 \over H_0^2}
       = {\Omega_{m0} (a_0/a)^3 + \Omega_{r0} (a_0/a)^4
       + \Omega_{\Lambda 0}
       + {1 \over 3} V / H_0^2
       \over 1 - {1 \over 6} \phi^{\prime 2}}, \quad
       {\dot H \over H_0^2}
       = - {3 \over 2} \Omega_{m0} \left( {a_0 \over a} \right)^3
       - 2 \Omega_{r0} \left( {a_0 \over a} \right)^4
       - {1 \over 2} {H^2 \over H_0^2} \phi^{\prime 2}.
   \label{BG-eqs-dimless}
\eea
During the QCD phase transition, which occurs deep in the radiation-dominated era, we simply use $H^2/H_0^2 = \Omega_{r0} (a_0/a)^4$, and similarly for $\dot H/H_0^2$. We set $a_0 \equiv 1$.

\subsubsection{WKB approximation}
                                  \label{App:WKB-BG}

As the evolution proceeds and $(m/H)^2 \gg 1$, the field $\phi$ oscillates too rapidly for its equation of motion to be solved accurately by direct numerical integration. To handle this regime, we switch from direct numerical integration to the WKB approximation at some epoch, as described below. In the rapidly oscillating stage with $\phi/f \ll 1$, in the radiation-dominated background, the equation of motion in Eq.\ (\ref{BG-eqs-dimless}) becomes
\bea
   \phi^{\prime\prime} + \phi^\prime
       + {m^2 \over H^2} \phi = 0.
\eea
For $(m/H)^2 \gg 1$, we have an approximate solution
\bea
   & & \phi = \overline c_1 \phi_1 + \overline c_2 \phi_2, \quad
       \phi_1 \equiv {1 \over \sqrt{m} a^{3/2}} 
       \sin{\overline \Omega}, \quad
       \phi_2 \equiv {1 \over \sqrt{m} a^{3/2}} 
       \cos{\overline \Omega}, \quad 
       \overline \Omega \equiv \int^N_{\overline N_{\rm tr}}
       {m \over H} d N, 
   \nonumber \\
   & & \phi_1^\prime = {1 \over \sqrt{m} a^{3/2}} \Big[
       {m \over H} \cos{\overline \Omega}
       - \Big({m^\prime \over 2 m} + {3 \over 2} \Big)
       \sin{\overline \Omega} \Big], \quad
       \phi_2^\prime = {1 \over \sqrt{m} a^{3/2}} \Big[
       - {m \over H} \sin{\overline \Omega}
       - \Big( {m^\prime \over 2 m} + {3 \over 2} \Big)
       \cos{\overline \Omega} \Big],
   \label{phi-sol-app}
\eea
where, without loss of generality, we choose the lower bound of integration to be $\overline N_{\rm tr}$, the transition epoch to the WKB regime. Alternatively, we have
\bea
   & & \phi = {\overline A \over \sqrt{m} a^{3/2}} 
       \cos( \overline \Omega + \overline B )
       = \overline c_1 \phi_1 + \overline c_2 \phi_2, \quad
       \phi_1 \equiv {\sin{\overline \Omega} \over \sqrt{m} a^{3/2}}, \quad
       \phi_2 \equiv {\cos{\overline \Omega} \over \sqrt{m} a^{3/2}},
       \quad
       \overline c_1 = - \overline A \sin{\overline B}, \quad
       \overline c_2 = \overline A \cos{\overline B},
   \nonumber \\
   & & 
       \phi^\prime = {\overline A \over \sqrt{m} a^{3/2}} \Big[
       - {m \over H} \sin{(\overline \Omega + \overline B)}
       - \Big( {m^\prime \over 2 m} + {3 \over 2} \Big)
       \cos{(\overline \Omega + \overline B)} \Big].
   \label{WKB-BG}
\eea 
The approximation is valid for
\bea
   \Big| \Big( {m^\prime \over 2 m} + {1 \over 2} \Big) 
       \Big( {m^\prime \over 2 m} + {3 \over 2} \Big)
       - \Big( {m^\prime \over 2 m} \Big)^\prime \Big|
       \ll {m^2 \over H^2}.
\eea

Coefficients $\overline c_1$ and $\overline c_2$ can be determined by matching $\phi$ and $\phi^\prime$ at the transition epoch. This leads to 
\bea
   \overline c_1 = {\phi \phi_2^\prime - \phi^\prime \phi_2 \over
       \phi_1 \phi_2^\prime - \phi_1^\prime \phi_2}, \quad
       \overline c_2 = - {\phi \phi_1^\prime - \phi^\prime \phi_1 \over
       \phi_1 \phi_2^\prime - \phi_1^\prime \phi_2},
\eea
where $\phi$ and $\phi^\prime$ are evaluated from the direct integration at the transition epoch. Alternatively, we have
\bea
   \overline B = - \tan^{-1}{\Big[ {H \over m} \Big(
       {\phi^\prime \over \phi}
       + {m^\prime \over 2m} + {3 \over 2} \Big) \Big]}
       = - \tan^{-1}{\Big( {\overline c_1 \over \overline c_2} \Big)},
       \quad
       \overline A = \sqrt{m} a^{3/2} {\phi \over \cos{\overline B}}
       = {\overline c_2 \over \cos{\overline B}}.
\eea
The two choices of $\overline B$ and $\overline B + \pi$ give the same $\phi$ and $\phi^\prime$.

The axion density and pressure in Eq.\ (\ref{BG-eqs-dimless}) give
\bea
   \Big( \begin{array}{c}
               \varrho_\phi \\
               p_\phi 
            \end{array} \Big)
       = {m \overline A^2 \over 2 a^3} 
       \Big[ 
            \Big( \begin{array}{c}
               1 \\
               -\cos{(2 \overline \Omega + 2 \overline B)}
            \end{array} \Big)
       + 2 {H \over m} \Big( {m^\prime \over 2 m} + {3 \over 2} \Big)
       \sin{(\overline \Omega + \overline B)} 
       \cos{(\overline \Omega + \overline B)}
       + {H^2 \over m^2} \Big( {m^\prime \over 2 m} + {3 \over 2} \Big)^2
       \cos^2{(\overline \Omega + \overline B)} \Big].
   \label{varrho-1}
\eea
To the leading order in $H/m$, the two rapidly oscillating terms in $\varrho_\phi$ combine to disappear. For $(H/m)^2 \ll 1$, we have
\bea
    \varrho_\phi \simeq {\overline A^2 \over 2 a^3} m, \quad
        p_\phi = - \cos{(2 \overline \Omega + 2 \overline B)} \varrho_\phi.
   \label{mu-WKB-approx}
\eea
Ignoring the rapid oscillation by time averaging, we have $p_\phi = 0$. Thus, as $m$ approaches $m_0$ near the end of the QCD phase transition, the axion behaves as CDM.

\subsection{Perturbation}
                                                 \label{Appendix-pert}

\subsubsection{Radiation-dominated approximation}

In the radiation-dominated era, $\delta_r$ in RCG (radiation-comoving gauge, $v_r \equiv 0$) has the growing mode solution \cite{Bardeen-1980}
\bea
   \delta_r = 4 C \Big( {\sin{x} \over x} - \cos{x} \Big), \quad
       \delta_r^\prime = 4 C \Big( 
       - {\sin{x} \over x} + \cos{x} + x \sin{x} \Big),
   \label{delta-r-sol-app}
\eea
where $x \equiv k \eta / \sqrt{3}$ with $\eta = 1/(aH)$. In the pure radiation era, Eqs.\ (\ref{eq6-r}) and (\ref{eq7-r}) give
\bea
   \alpha = - {1 \over 4} \delta_r, \quad
       {\kappa \over H} = {3 \over 4} ( \delta_r^\prime - \delta_r ).
   \label{alpha-sol-app}
\eea
In this approximation, we only need to solve the equation for $\delta \phi$. Equation (\ref{EOM-pert}) gives
\bea
   \delta \phi^{\prime\prime}
       + \Big( 3 + {\dot H \over H^2} \Big)
       \delta \phi^\prime
       + \Big[ \Big( {k \over aH} \Big)^2
       + {V_{,\phi\phi} \over H^2} \Big] \delta \phi
       = {1 \over 2} \phi^\prime \delta_r^\prime
       + \Big( {V_{,\phi} \over 2 H^2}
       - {V_{,\phi \varrho_r} \varrho_r \over H^2} \Big) \delta_r \equiv S.
   \label{ddot-delta-phi-app}
\eea
The relative axion density perturbation $\delta_\phi$ can be constructed using
\bea
   \delta_\phi \equiv {\delta \varrho_\phi \over \varrho_\phi}
       = {\phi^\prime \delta \phi^\prime
       + {V_{,\phi} \over H^2} \delta \phi
       - \phi^{\prime 2} \alpha
       + {V_{,\varrho_r} \varrho_r \over H^2} \delta_r
       \over {1 \over 2} \phi^{\prime 2}
       + {V \over H^2}}.
   \label{delta_phi}
\eea

For comparison, we present the evolution equation for the CDM perturbation in a radiation-dominated environment in RCG. Using the energy and momentum conservation equations, Eqs.\ (\ref{eq6}) and (\ref{eq7}), for the CDM
\bea
   \delta \dot \varrho_c + 3 H \delta \varrho_c 
       = \varrho_c \Big( \kappa - 3 H \alpha 
       + {\Delta \over a} v_c \Big), \quad
       {1 \over a} ( a v_c )^{\displaystyle{\cdot}} = {1 \over a} \alpha,
\eea
and Eq.\ (\ref{eq4}), we have
\bea
   \ddot \delta_c + 2 H \dot \delta_c - 4 \pi G \varrho_c \delta_c
       = {3 \over 4} H \dot \delta_r + {3 \over 2} H^2 \delta_r,
\eea
with $\delta_r$ given in Eq.\ (\ref{delta-r-sol-app}). Ignoring the gravity term of CDM (the third term), we have
\bea
   \delta_c^{\prime\prime} = {3 \over 4} \delta_r^\prime
       + {3 \over 2} \delta_r.
   \label{delta-eq-CDM}
\eea

\subsubsection{Initial conditions}

We derive initial conditions for the background and perturbed scalar field in the presence of the dominant radiation component. At the initial epoch, we assume the large-scale (super-horizon) limit, before QCD phase transition (thus $\varrho_r \gg \varrho_{rc}$), and a radiation-dominated background. The equations become
\bea
   \phi^{\prime\prime} + \phi^\prime 
       = - {V_{,\phi} \over H^2}
       \equiv \overline S, \quad
       \delta \phi^{\prime\prime} + \delta \phi^\prime = S
       - \Big( {m^2 \over H^2} + {k^2 \over a^2 H^2} \Big)
       \delta \phi 
       \equiv \widetilde S.
\eea
We introduce $\overline S$ as a source term by assuming that $\phi$ remains approximately constant during the initial epoch, so that $\overline S$ is independent of $\phi$. Similarly, we absorb the $\delta \phi$ term into the source term when $\delta \phi$ remains approximately constant during the initial epoch. The general solutions are
\bea
   \phi = \int^N \overline S d N 
       - {1 \over a} \int^N a \overline S d N, \quad
       \delta \phi = \int^N \widetilde S d N 
       - {1 \over a} \int^N a \widetilde S d N,
   \label{solutions}
\eea
where the homogeneous solutions are absorbed to the lower bounds of integrations. Using Eq.\ (\ref{V-X}) we have
\bea
   & & \overline S = - {V_{,\phi} \over H^2}
       \simeq - {1 \over H^2} f 
       {m_0^2 \over 1 + (\varrho_r/\varrho_{rc})^q} 
       \sin{(\phi/f)} \propto \varrho_r^{-1-q} 
       \propto a^{4 + 4 q},
   \nonumber \\
   & & 
       \widetilde S = {3 + 14 q + 8 q^2 \over 2 (5 + 4q)} 
       {V_{,\phi} \over H^2} \delta_r 
       - \overline c_d e^{-N} \delta_r
       - {\omega^2 \over H^2} \delta \phi_i, \quad
       \delta_r = {4 \over 3} C x^2 \propto a^2.
\eea
From Eq.\ (\ref{solutions}), we have general solutions
\bea
   & & \phi = \overline c_g + \overline c_d e^{-N}
      -  {1 \over (4 + 4q) (5 + 4q)} {V_{,\phi} \over H^2}, 
   \nonumber \\
   & &
       \delta \phi = c_g + c_d e^{-N}
       - {1 \over 2} \overline c_d e^{-N} \delta_r
       +{3 + 14 q + 8 q^2 \over
       2 (6 + 4 q) (7 + 4q) (5+4q)} {V_{,\phi} \over H^2} \delta_r
       -{1 \over 6} \Big( {k \over aH} \Big)^2 \delta \phi_i
       - {1 \over 20} {m^2 \over H^2} \delta \phi_i, 
   \label{ICs}
\eea
where $\overline c_i$ and $c_i$ are constant coefficients of homogeneous solutions. The last two terms in $\delta \phi$ are proportional to $a^2$ and $a^4$, respectively, with $\delta \phi_i = c_g$; in the early era the first term is dominant. These expressions encompass the general initial conditions. For the background, $\phi$ is provided by the initial position of the field like the bottom ($\phi/f \ll 1$) or the hilltop ($\phi/f \simeq \pi$) conditions. We may ignore the decaying modes by choosing $\overline c_d = 0 = c_d$. For the radiation-supported initial conditions, we can choose the terms with $V_{,\phi}/H^2$ in $\phi^\prime$, $\delta \phi$ and $\delta \phi^\prime$. In this radiation-dominated regime, the initial conditions are completely determined except for $\overline c_g$, which is fixed by the initial position of $\phi$ in the potential.  

For the CDM, as the initial condition, in the large-scale limit, using $\delta_r = {4 \over 3} C x^2$, Eq.\ (\ref{delta-eq-CDM}) gives
\bea
   \delta_c = C_g N + C_d + {3 \over 4} \delta_r
\eea
where $C_g$ and $C_d$ are constant coefficients of the homogeneous solution. Thus, for the CDM perturbation in the radiation-dominated era, we take $\delta_c = {3 \over 4} \delta_r$ as the initial condition.

\subsubsection{WKB approximation}
                                      \label{Appendix-WKB-pert}

Because both the background and perturbed scalar fields undergo rapid oscillations, direct numerical integration of Eq.\ (\ref{ddot-delta-phi-app}) eventually becomes impractical before the axion evolution can be approximated as CDM by averaging over the oscillations. Similarly to the background evolution, we use the WKB approximation \cite{Sikivie-Xue-2022}.

In the radiation-dominated era, with $\phi/f \ll 1$, Eq.\ (\ref{ddot-delta-phi-app}) becomes
\bea
   \delta \phi^{\prime\prime}
       + \delta \phi^\prime
       + {\omega^2\over H^2} \delta \phi
       = S, \quad
       \omega \equiv \sqrt{V_{,\phi\phi} + {k^2 \over a^2}}.
   \label{WKB-pert-eq}
\eea
In the rapidly oscillating phase, we have $\phi/f \ll 1$, and thus $V_{,\phi\phi} = m^2$. In the WKB regime we can use Eq.\ (\ref{phi-sol-app}) for background, Eqs.\ (\ref{delta-r-sol-app}) and (\ref{alpha-sol-app}) for radiation and metric perturbations. For $(\omega/H)^2 \gg 1$, as the homogeneous solutions of Eq.\ (\ref{WKB-pert-eq}), ignoring the right-hand side, we have an approximate solution
\bea
   & & \delta \phi_{\rm H} 
       = c_1 \delta \phi_1 + c_2 \delta \phi_2, \quad
       \delta \phi_1 \equiv {\sin{\Omega} \over \sqrt{\omega} a^{3/2}} 
       , \quad
       \delta \phi_2 \equiv {\cos{\Omega} \over \sqrt{\omega} a^{3/2}} 
       , \quad
       \Omega \equiv \int^N_{N_{\rm tr}}
       {\omega \over H} d N, 
   \nonumber \\
   & & \delta \phi_1^\prime = {1 \over \sqrt{\omega} a^{3/2}} 
       \Big[ {\omega \over H} \cos{\Omega} 
       - \Big( {\omega^\prime \over 2 \omega} + {3 \over 2} \Big)
       \sin{\Omega} \Big], \quad
       \delta \phi_2^\prime = {1 \over \sqrt{\omega} a^{3/2}} 
       \Big[ -  {\omega \over H} \sin{\Omega}
       - \Big( {\omega^\prime \over 2 \omega} + {3 \over 2} \Big)
       \cos{\Omega} \Big],
   \label{WKB-pert}
\eea
where, without loss of generality, we choose the lower bound of integration to be $N_{\rm tr}$, the transition epoch to the WKB regime.
Alternatively, we have 
\bea
   & & \delta \phi_{\rm H} = {A \over \sqrt{\omega} a^{3/2}} 
       \cos( \Omega + B ), \quad
       c_1 = - A \sin{B}, \quad
       c_2 = A \cos{B},
   \nonumber \\
   & & 
       \delta \phi^\prime_{\rm H} 
       = {A \over \sqrt{\omega} a^{3/2}} \Big[
       - {\omega \over H} \sin{(\Omega + B)}
       - \Big( {\omega^\prime \over 2 \omega} + {3 \over 2} \Big)
       \cos{(\Omega + B)} \Big].
   \label{WKB-pert-2}
\eea 
This homogeneous solution is valid for
\bea
   & & 
       \Big| \Big( {\omega^\prime \over 2 \omega} + {1 \over 2} \Big) 
       \Big( {\omega^\prime \over 2 \omega} + {3 \over 2} \Big)
       - \Big( {\omega^\prime \over 2 \omega} \Big)^\prime \Big|
       \ll {\omega^2 \over H^2}.
\eea
The general solutions are
\bea
   & & \delta \phi = \delta \phi_{\rm H}
      + \delta \phi_1 \int^N_{N_{\rm tr}} S H a^3 
       \delta \phi_2 d N
       - \delta \phi_2 \int^N_{N_{\rm tr}} S H a^3
       \delta \phi_1 d N, 
   \nonumber \\
   & & 
       \delta \phi^\prime 
       = \delta \phi_{\rm H}^\prime
       + \delta \phi_1^\prime
       \int^N_{N_{\rm tr}} S H a^3 \delta \phi_2 d N
       - \delta \phi_2^\prime \int^N_{N_{\rm tr}} S H a^3
       \delta \phi_1 d N,
   \label{general-solution-app}
\eea
where, without loss of generality, we choose the lower bound of integrations as $N_{\rm tr}$; their effects can be absorbed into $c_1$ and $c_2$, or equivalently into $A$ and $B$.

Coefficients $c_1$ and $c_2$ can be determined by matching $\delta \phi$ and $\delta \phi^\prime$ at the transition epoch. This leads to 
\bea
   c_1 = {\delta \phi \delta \phi_2^\prime 
       - \delta \phi^\prime \delta \phi_2 \over
       \delta \phi_1 \delta \phi_2^\prime 
       - \delta \phi_1^\prime \delta \phi_2}, \quad
       c_2 = - {\delta \phi \delta \phi_1^\prime
       - \delta \phi^\prime \delta \phi_1 \over
       \delta \phi_1 \delta \phi_2^\prime
       - \delta \phi_1^\prime \delta \phi_2},
\eea
where $\delta \phi$ and $\delta \phi^\prime$ are evaluated from the direct integration at the transition epoch. Alternatively, we have
\bea
   B = - \tan^{-1}{\Big[ {H \over \omega} \Big(
       {\delta \phi^\prime \over \delta \phi}
       + {\omega^\prime \over 2\omega} + {3 \over 2} \Big) \Big]}
       = - \tan^{-1}{\Big({c_1 \over c_2} \Big)}, \quad
       A = \sqrt{\omega} a^{3/2} {\delta \phi \over \cos{B}}
       = {c_2 \over \cos{B}}.
\eea
The two alternative choices of $B$, $B$ or $B + \pi$, give the same $\delta \phi$ and $\delta \phi^\prime$.

The density and pressure perturbations become
\bea
   & & \Big( \begin{array}{c}
               \delta \varrho_\phi \\
               \delta p_\phi 
            \end{array} \Big)
      = H^2 \phi^\prime \delta \phi^\prime
      \pm m^2 \phi \delta \phi
      - H^2 \phi^{\prime 2} \alpha
      \pm V_{,\varrho_r} \varrho_r \delta_r
   \nonumber \\
   & & \qquad
      = {m \overline A^2 \over 2 a^3} {2 \over \overline A}
      \sqrt{m \over \omega}
      \bigg\{ A \Big[ {\omega \over m} 
      \sin{(\overline \Omega + \overline B)}
      \sin{(\Omega + B)}
      \pm \cos{(\overline \Omega + \overline B)}
      \cos{(\Omega + B)}
   \nonumber \\
   & & \qquad
      + {H \over m} 
      \Big( {\omega^\prime \over 2 \omega} + {3 \over 2} \Big)
      \sin{(\overline \Omega + \overline B)}
      \cos{(\Omega + B)}
      + {H \over m} {\omega \over m} 
      \Big( {m^\prime \over 2 m} + {3 \over 2} \Big)
      \cos{(\overline \Omega + \overline B)}
      \sin{(\Omega + B)}
   \nonumber \\
   & & \qquad
      + {H^2 \over m^2} \Big( {m^\prime \over 2 m} + {3 \over 2} \Big)
      \Big( {\omega^\prime \over 2 \omega} + {3 \over 2} \Big)
      \cos{(\overline \Omega + \overline B)}
      \cos{(\Omega + B)} \Big]
   \nonumber \\
   & & \qquad
      + \Big[ - {\omega \over m} 
      \sin{(\overline \Omega + \overline B)} \cos{\Omega}
      \pm \cos{(\overline \Omega + \overline B)} \sin{\Omega}
   \nonumber \\
   & & \qquad
      + {H \over m} \Big( {\omega^\prime \over 2 \omega} 
      + {3 \over 2} \Big) \sin{(\overline \Omega + \overline B)}
      \sin{\Omega}
      - {H \over m} {\omega \over m} \Big( {m^\prime \over 2 m} 
      + {3 \over 2} \Big) \cos{(\overline \Omega + \overline B)}
      \cos{\Omega}
   \nonumber \\
   & & \qquad
      + {H^2 \over m^2} \Big( {m^\prime \over 2 m} + {3 \over 2} \Big) 
      \Big( {\omega^\prime \over 2 \omega} + {3 \over 2} \Big) 
      \cos{(\overline \Omega + \overline B)}
      \sin{\Omega} \Big]
      \int^N_{N_{\rm tr}} S H a^3 \delta \phi_2 d N
    \nonumber \\
    & & \qquad
      - \Big[ {\omega \over m} 
      \sin{(\overline \Omega + \overline B)} \sin{\Omega}
      \pm \cos{(\overline \Omega + \overline B)} \cos{\Omega}
   \nonumber \\
   & & \qquad
      + {H \over m} \Big( {\omega^\prime \over 2 \omega} 
      + {3 \over 2} \Big) \sin{(\overline \Omega + \overline B)}
      \cos{\Omega}
      + {H \over m} {\omega \over m} \Big( {m^\prime \over 2 m} 
      + {3 \over 2} \Big) \cos{(\overline \Omega + \overline B)}
      \sin{\Omega}
   \nonumber \\
   & & \qquad
      + {H^2 \over m^2} \Big( {m^\prime \over 2 m} + {3 \over 2} \Big) 
      \Big( {\omega^\prime \over 2 \omega} + {3 \over 2} \Big) 
      \cos{(\overline \Omega + \overline B)}
      \cos{\Omega} \Big]
      \int^N_{N_{\rm tr}} S H a^3 \delta \phi_1 d N 
      \bigg\}
   \nonumber \\
   & & \qquad
      + {m \overline A^2 \over 2 a^3} \bigg\{
      {1 \over 2} \Big[ \sin{(\overline \Omega + \overline B)}
      + {H \over m} \Big( {m^\prime \over 2m} + {3 \over 2} \Big)
      \cos{(\overline \Omega + \overline B)} \Big]^2 
      \pm \cos^2{(\overline \Omega + \overline B)}
      {X_{,\varrho_r} \over X} \varrho_r 
      \bigg\} \delta_r.
   \label{delta-phi-WKB1}
\eea
The above formula is exact in the WKB regime under our ansatz in Eqs.\ (\ref{WKB-BG}) and (\ref{general-solution-app}). Although we used these exact forms of the WKB approximation in the numerical computation of the power spectrum of axion density perturbation, it is convenient to proceed with the analytic calculation a bit further. The rapid oscillation of the first two terms in the numerator of $\delta_\phi$ in Eq.\ (\ref{delta_phi}) should exactly add up to yield a smooth result; a similar cancellation of the oscillatory terms occurs in the background axion density, as shown in Eq.\ (\ref{mu-WKB-approx}). Using 
\bea
   S = {\overline A \over \sqrt{m} a^{3/2}} 
       \big[ S_1 \cos{(\overline \Omega + \overline B)}
       - S_2 \sin{(\overline \Omega + \overline B)} \big], \quad
       S_1 \equiv {m^2 \over H^2} \Big( {1 \over 2}
       - {X_{,\varrho_r} \over X} \varrho_r \Big) \delta_r
       - {1 \over 2} \Big( {m^\prime \over 2m}
       + {3 \over 2} \Big) \delta_r^\prime, \quad
       S_2 \equiv {1 \over 2} {m \over H} \delta_r^\prime,
\eea
and neglecting rapidly oscillating terms such as $\sin{(\Omega +\overline \Omega + ...)}$ and $\cos{(\Omega +\overline \Omega + ...)}$ in the fluid quantities, we have
\bea
   & & \delta \varrho_{\phi} \simeq 
      {m \overline A^2 \over 2 a^3} {A \over \overline A}
       \sqrt{m \over \omega} \bigg\{ \Big( 1 + {\omega \over m} \Big)
       \cos{(\Delta \Omega + \Delta B)}
       - {H \over m} \Big[ {\omega^\prime \over 2 \omega}
       + {3 \over 2} - {\omega \over m} 
       \Big( {m^\prime \over 2 m} + {3 \over 2} \Big) \Big]
       \sin{(\Delta \Omega + \Delta B)} 
   \nonumber \\
   & & \qquad
       + {H^2 \over m^2} 
       \Big( {m^\prime \over 2 m} + {3 \over 2} \Big)
       \Big( {\omega^\prime \over 2 \omega} + {3 \over 2} \Big)
       \cos{(\Delta \Omega + \Delta B)} \bigg\} 
       + {m \overline A^2 \over 2 a^3} {1 \over 2} 
       \Big[ {1 \over 2} + {X_{,\varrho_r} \over X} \varrho_r
       + {H^2 \over m^2} \Big( {m^\prime \over 2 m} + {3 \over 2}
       \Big) \Big] \delta_r
   \nonumber \\
   & & \qquad
       + {m \overline A^2 \over 2 a^3} {1 \over 2} 
       \sqrt{m \over \omega}
       \bigg\{ \Big( 1 + {\omega \over m} \Big)
       \int_{N_{\rm tr}}^N {H \over \sqrt{m\omega}} 
       \big[ - S_1 \sin{(\Delta \widetilde \Omega - \Delta \Omega)}
       + S_2 \cos{(\Delta \widetilde \Omega - \Delta \Omega)}
       \big] d \widetilde N
   \nonumber \\
   & & \qquad
       + {H \over m} \Big[ {\omega^\prime \over 2 \omega} + {3 \over 2}
       - {\omega \over m} \Big({m^\prime \over 2 m} + {3 \over 2} \Big)
       \Big] \int_{N_{\rm tr}}^N {H \over \sqrt{m\omega}} 
       \big[ S_1 \cos{(\Delta \widetilde \Omega - \Delta \Omega)}
       + S_2 \sin{(\Delta \widetilde \Omega - \Delta \Omega)}
       \big] d \widetilde N
   \nonumber \\
   & & \qquad
       + {H^2 \over m^2} 
       \Big({m^\prime \over 2 m} + {3 \over 2} \Big) 
       \Big( {\omega^\prime \over 2 \omega} + {3 \over 2} \Big)
       \int_{N_{\rm tr}}^N {H \over \sqrt{m\omega}} 
       \big[ - S_1 \sin{(\Delta \widetilde \Omega - \Delta \Omega)}
       + S_2 \cos{(\Delta \widetilde \Omega - \Delta \Omega)}
       \big] d \widetilde N \bigg\},
   \label{delta-phi-WKB2}
\eea
where $\Delta \Omega \equiv \Omega - \overline \Omega$, $\Delta \widetilde \Omega \equiv \widetilde \Omega - \widetilde {\overline \Omega}$ and $\Delta B \equiv B - \overline B$. For $\omega \simeq m$, we have $\Delta \Omega \simeq 0 \simeq \Delta \widetilde \Omega$, and thus
\bea
   & & \delta \varrho_{\phi}
       \simeq {m \overline A^2 \over 2 a^{3}} \bigg\{
       \Big[ 1 + {1 \over 2} {H^2 \over m^2} 
       \Big({m^\prime \over 2 m} + {3 \over 2} \Big)^2 \Big]
       \Big[ 2 {A \over \overline A} \cos{(\Delta \Omega 
       +\Delta B)}
       + {1 \over 2} [ \delta_r - \delta_r(N_{\rm tr}) ] \Big]
   \nonumber \\
   & & \qquad
       + {1 \over 2} \Big[ {1 \over 2} + {X_{,\varrho_r} \over X} \varrho_r
       + {H^2 \over m^2} \Big( {m^\prime \over 2 m} + {3 \over 2}
       \Big) \Big] \delta_r \bigg\},
   \label{delta-phi-WKB3}
\eea
where we kept $\Delta \Omega$ which makes the approximation more accurate. Neglecting higher-order terms in $H^2/m^2$, we obtain
\bea
   \delta_{\phi}
       \simeq 2 {A \over \overline A} \cos{(\Delta \Omega 
       + \Delta B)} 
       + {1 \over 2} [ \delta_r - \delta_r(N_{\rm tr}) ]
       + {1 \over 2} \Big( {1 \over 2} + {X_{,\varrho_r} \over X} \varrho_r
       \Big) \delta_r.
   \label{delta-phi-WKB4}
\eea
Thus, at later times, when $X$ has settled to unity, we have $\delta_\phi = {3 \over 4} \delta_r$ as expected for the adiabatic mode of CDM in the radiation era. In the later ordinary axion stage with $X = {\rm constant}$, still in the radiation-dominated era, we have
\bea
   \delta_\phi \simeq 2 {A \over \overline A} 
      \cos{(\Delta \Omega + \Delta B)}
      + {3 \over 4} \delta_r - {1 \over 2} \delta_r (t_{\rm tr}),
\eea
where $t - t_{\rm tr} = \int_{N_{\rm tr}}^N (1/H) dN$, $t_{\rm tr} /t = e^{2(N_{\rm tr} -N)}$, and $\delta_r (t_{\rm tr}) = \delta_r (N_{\rm tr})$.

\end{widetext}
\end{document}